\documentclass[letterpaper,11pt]{article}

\usepackage{fullpage}
\usepackage{newtxtext}
\usepackage{microtype}
\usepackage{enumitem}
\usepackage[dvipsnames]{xcolor}
\usepackage{breakcites}
\usepackage{booktabs}

\RequirePackage{mathtools}

\usepackage{amsfonts,amssymb}

\renewcommand{\cal}[1]{\mathcal{#1}}

\makeatletter
\renewcommand{\pod}[1]{\mathchoice
  {\allowbreak \if@display \mkern 18mu\else \mkern 8mu\fi (#1)}
  {\allowbreak \if@display \mkern 18mu\else \mkern 8mu\fi (#1)}
  {\mkern4mu(#1)}
  {\mkern4mu(#1)}
}
\makeatletter
\ifcsname NewCommandCopy\endcsname
\NewCommandCopy\@@pmod\pmod
\else%
\let\@@pmod\pmod
\fi%
\DeclareRobustCommand{\pmod}{\@ifstar\@pmods\@@pmod}
\def\@pmods#1{\mkern4mu({\operator@font mod}\mkern 6mu#1)}
\makeatother
\RequirePackage{mathtools}

\newcommand{\C}{\ensuremath{\mathbb{C}}}

\newcommand{\F}{\ensuremath{\mathbb{F}}}

\newcommand{\R}{\ensuremath{\mathbb{R}}}

\newcommand{\Z}{\ensuremath{\mathbb{Z}}}

\newcommand{\Zq}{\ensuremath{\Z_q}}

\RequirePackage{mathtools}

\DeclarePairedDelimiterX\inprod[2]{\langle}{\rangle}{#1,#2}
\DeclarePairedDelimiter\abs{\lvert}{\rvert}

\DeclarePairedDelimiter\set{\{}{\}}
\DeclarePairedDelimiter\parens{(}{)}

\DeclarePairedDelimiter\bracks{[}{]}

\DeclarePairedDelimiter\floor{\lfloor}{\rfloor}
\DeclarePairedDelimiter\ceil{\lceil}{\rceil}
\DeclarePairedDelimiter\length{\lVert}{\rVert}
\DeclarePairedDelimiter\norm{\lVert}{\rVert}

\DeclarePairedDelimiterX\braket[2]{\langle}{\rangle}{#1 \delimsize\vert #2}

\RequirePackage[pagebackref=true]{hyperref}

\hypersetup{
        hypertexnames=false,
        breaklinks=true,
        colorlinks=true,
        linkcolor=blue,
        filecolor=blue,
        citecolor=blue,
        urlcolor=blue,
}

\usepackage[all]{hypcap}

\RequirePackage{zref-clever}
\zcsetup{
        cap=true, %
        nameinlink,
}

\newcommand{\cref}[1]{\zcref{#1}}
\newcommand{\Cref}[1]{\zcref[S]{#1}}

\makeatletter
\@ifpackageloaded{backref}{%
\renewcommand*{\backref}[1]{}
\renewcommand*{\backrefalt}[4]{{%
    \ifcase #1 Not cited.%
    \or Page~#2.%
    \else Pages #2.%
    \fi%
  }}}{}
\makeatother
\RequirePackage{mathtools}

\newcommand{\matB}{\ensuremath{\mathbf{B}}}

\newcommand{\vecb}{\ensuremath{\mathbf{b}}}
\newcommand{\vecc}{\ensuremath{\mathbf{c}}}

\newcommand{\vece}{\ensuremath{\mathbf{e}}}

\newcommand{\vecr}{\ensuremath{\mathbf{r}}}

\newcommand{\vect}{\ensuremath{\mathbf{t}}}
\newcommand{\vecu}{\ensuremath{\mathbf{u}}}
\newcommand{\vecv}{\ensuremath{\mathbf{v}}}
\newcommand{\vecw}{\ensuremath{\mathbf{w}}}
\newcommand{\vecx}{\ensuremath{\mathbf{x}}}
\newcommand{\vecy}{\ensuremath{\mathbf{y}}}

\newcommand{\veczero}{\ensuremath{\mathbf{0}}}

\usepackage[amsmath,amsthm,thmmarks,hyperref]{ntheorem}

\RequirePackage{thmtools}           %

\declaretheorem[numberwithin=section]{theorem}

\declaretheorem[sibling=theorem,style=plain]
        {lemma,corollary,proposition,claim,fact,maintheorem}

\declaretheorem[sibling=theorem,style=definition]
        {definition,conjecture,construction,assumption}

\declaretheorem[sibling=theorem,style=remark]
        {remark,example,note}

\zcRefTypeSetup{claim}{Name-sg=Claim,Name-pl=Claims,name-sg=claim,name-pl=claims}
\zcRefTypeSetup{fact}{Name-sg=Fact,Name-pl=Facts,name-sg=fact,name-pl=facts}
\zcRefTypeSetup{maintheorem}{Name-sg=Main Theorem,Name-pl=Main
Theorems,name-sg=main theorem,name-pl=main theorems}
\zcRefTypeSetup{conjecture}{Name-sg=Conjecture,Name-pl=Conjectures,name-sg=conjecture,name-pl=conjectures}
\zcRefTypeSetup{construction}{Name-sg=Construction,Name-pl=Constructions,name-sg=construction,name-pl=constructions}
\zcRefTypeSetup{assumption}{Name-sg=Assumption,Name-pl=Assumptions,name-sg=assumption,name-pl=assumptions}

\numberwithin{equation}{section}
\RequirePackage{mathtools}
\RequirePackage{xspace}

\newcommand{\bit}{\ensuremath{\set{0,1}}}

\DeclareMathOperator{\poly}{poly}

\DeclareMathOperator*{\E}{\mathbb{E}}

\RequirePackage{mathtools}

\newcommand{\lat}{\mathcal{L}}

\DeclareMathOperator{\spn}{span}

\makeatletter
\newcommand*{\newreptext}[1]{%
  \begingroup %
  \csname @safe@actives@true\endcsname
  \expandafter\endgroup
  \expandafter\newcommand\csname reptext@#1\endcsname
}

\newcommand*{\reptext}[1]{%
  \begingroup
  \csname @safe@actives@true\endcsname %
  \@ifundefined{reptext@#1}{%
    \@latex@error{\string\reptext{#1} is undefined}\@ehc
    \endgroup
    \textbf{??}%
  }{%
    \endgroup
    \@nameuse{reptext@#1}%
  }%
}
\makeatother

\newcommand{\eps}{\varepsilon}

\newcommand{\code}{\cal{C}}

\newcommand{\grs}{\text{GRS}}

\DeclareMathOperator{\corr}{corr}
\DeclareMathOperator*{\Avg}{Avg}

\newcommand{\vecalpha}{\boldsymbol{\alpha}}

\newcommand{\defn}{\vcentcolon=}

\usepackage{derivative}

\title{List Decoding Reed--Solomon Codes \\
  in the Lee, Euclidean, and Other Metrics}

\author{Chris Peikert\thanks{University of Michigan, \texttt{cpeikert@umich.edu}.}
  \and Alexandra Veliche Hostetler\thanks{University of Michigan, \texttt{aveliche@umich.edu}.}}

\begin{document}

\maketitle

\begin{abstract}
Reed--Solomon error-correcting codes are ubiquitous across computer science and information theory, with applications in cryptography, computational complexity, communication and storage systems, and more.
Most works on efficient error correction for these codes, like the celebrated Berlekamp--Welch unique decoder and the (Guruswami--)Sudan list decoders, are focused on measuring error in the Hamming metric, which simply counts the number of corrupted codeword symbols.
However, for some applications, other metrics that depend on the specific values of the errors may be more appropriate.

This work gives a polynomial-time algorithm that list decodes (generalized) Reed--Solomon codes over prime fields in~$\ell_p$ (semi)metrics, for any $0 < p \leq 2$.
Compared to prior algorithms for the Lee~($\ell_1$) and Euclidean~($\ell_2$) metrics, ours decodes to arbitrarily large distances (for correspondingly small rates), and has better distance-rate tradeoffs for all decoding distances above some moderate thresholds.
We also prove lower bounds on the~$\ell_{1}$ and~$\ell_{2}$ minimum distances of a certain natural subclass of GRS codes, which establishes that our list decoder is actually a \emph{unique} decoder for many parameters of interest.
Finally, we analyze our algorithm's performance under \emph{random} Laplacian and Gaussian errors, and show that it supports even larger rates than for corresponding amounts of worst-case error in~$\ell_{1}$ and~$\ell_{2}$ (respectively).
\end{abstract}

\section{Introduction}%
\label{sec:introduction}

Reed--Solomon codes~\cite{ReedSolomon} are among the most widely used families of error-correcting codes, with applications across computer and communication sciences.
Their many virtues include: a very simple definition; the largest possible minimum distance as a function of rate; and efficient decodability from errors, via either \emph{unique} decoding up to half the minimum distance (see, e.g.,~\cite[Section~12.1]{GRS:_essential_coding_theory}), or \emph{list} decoding up to the larger Johnson bound, via the celebrated works of Sudan~\cite{DBLP:journals/jc/Sudan97} and Guruswami--Sudan~\cite{DBLP:journals/tit/GuruswamiS99} (see also~\cite[Section~12.2]{GRS:_essential_coding_theory}).

List decoding~\cite{elias-zero-error,wozencraft} is the task of finding all codewords that are within some desired distance of a (potentially corrupted) received word.
When this radius is more than half the code's minimum distance, there can potentially be more than one codeword within range (hence the name ``list decoding'').
Despite this non-uniqueness, list decoding can suffice for many purposes (e.g., finding a nearest codeword within range), and indeed, it has found numerous applications.

Most work on decoding Reed--Solomon codes has measured errors in the \emph{Hamming} metric, which simply counts the \emph{number} of corrupted codeword symbols (regardless of how they are corrupted).
However, there are many other natural metrics that depend on the specific \emph{values} of the errors.
Such metrics can be more appropriate for settings where introducing a ``large'' error at a coordinate is more costly than a ``small'' error, or where the communication channel might add some nonzero error to every coordinate.
An example is a channel that adds error according to a Gaussian or other fairly concentrated distribution.
When the code alphabet is~$\Zq$ (the integers modulo~$q$)---in particular, a prime field~$\F_{q}$---one metric of frequent study is the Lee metric, which is merely the~$\ell_{1}$ norm $\norm{\vecx}_{1} = \sum_{i} \abs{x_{i}}$ after lifting~$\Zq$ to its distinguished representatives in $[-q/2,q/2)$.
Other natural, analogously defined choices include the Euclidean~($\ell_{2}$) or other~$\ell_{p}$ metrics.

We know of only a few prior works on efficiently decoding Reed--Solomon codes in metrics other than Hamming.
For the Lee~($\ell_{1}$) metric, Roth and Siegel~\cite{DBLP:journals/tit/RothS94} gave an algorithm that uniquely decodes up to half of (a lower bound on) the minimum distance; their algorithm works for certain subclasses of (generalized) Reed--Solomon and BCH codes.
In addition, Wu, Kuijper, and Udaya~\cite{wu03:_lee_bch_reed_solomon} gave a list-decoding algorithm for~$\ell_{1}$, built around Guruswami--Sudan~\cite{DBLP:journals/tit/GuruswamiS99}, that decodes to larger distances than in~\cite{DBLP:journals/tit/RothS94} for all small enough rates.
Finally, for the Euclidean~($\ell_{2}$) metric, Mook and Peikert~\cite{DBLP:journals/tit/MookP22} gave a list-decoding algorithm that also uses~\cite{DBLP:journals/tit/GuruswamiS99} as a black box.

\subsection{Contributions}%
\label{sec:contributions}

\begin{figure}[t]
  \centering
  \includegraphics[width=0.48\textwidth]{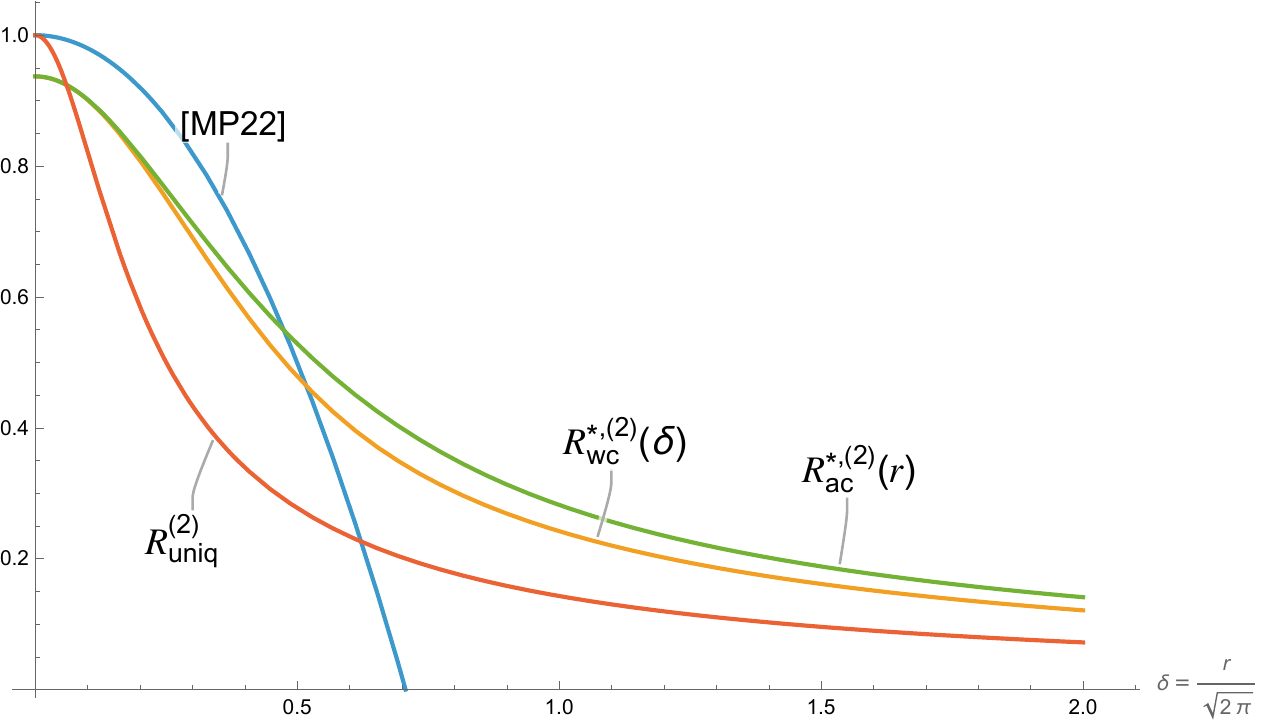}
  \includegraphics[width=0.48\textwidth]{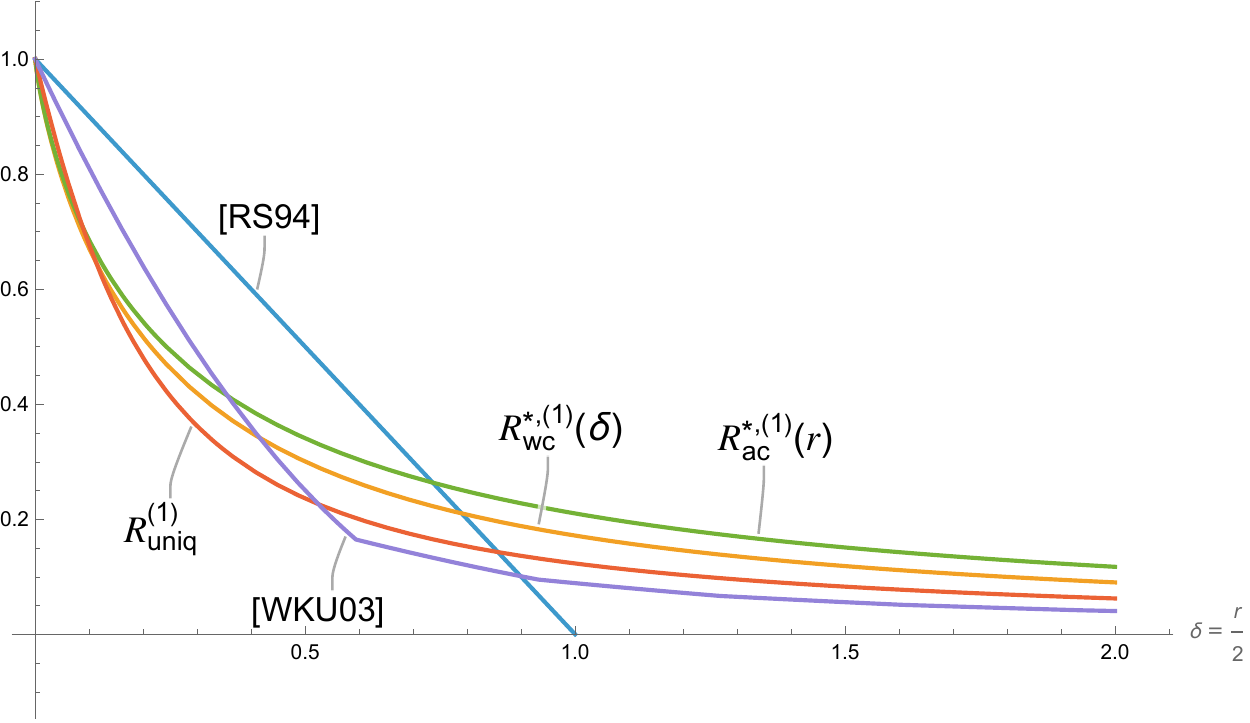}

  \caption{Plots of the adjusted rate~$R^{*,(p)}$, as a function of the~$\ell_{p}$ relative decoding distance~$\delta = d/n^{1/p}$ or corresponding channel error width $r = p^{1/p} \cdot c_{p} \cdot \delta$, for which our algorithm can list decode prime-field GRS codes in the worst case (wc) or average case (ac), respectively, for $p=2$ (left) and $p=1$ (right).
    (For simplicity, these plots assume a field size $q \gg \delta, r$.)
    For comparison, also shown are the corresponding functions from the prior work on decoding GRS codes in these metrics: \cite{DBLP:journals/tit/MookP22} is for list decoding in~$\ell_{2}$, and~\cite{DBLP:journals/tit/RothS94,wu03:_lee_bch_reed_solomon} are respectively for \emph{unique} and \emph{list} decoding in the~$\ell_{1}$ (Lee) metric, but only for \emph{discrete} (integer) error.
    Also shown are rate bounds $R^{(p)}_{\text{uniq}}$ for which decoding to~$\ell_{p}$ relative distance~$\delta$ is guaranteed to yield a \emph{unique} codeword, for a certain natural subclass of GRS codes.
    (See \cref{lem:ell2-min-dist,lem:ell1-min-dist} and the discussions thereafter.)}%
  \label{fig:plots}
\end{figure}

This work gives a polynomial-time algorithm that list decodes any generalized Reed--Solomon~(GRS) code over a prime field in the~$\ell_p$ (semi)metric for any $0 < p \leq 2$; in particular, this includes the Lee~($\ell_{1}$) and Euclidean~($\ell_{2}$) metrics.\footnote{A semimetric is just a metric that does not necessarily satisfy the triangle inequality (which we will not need).}
Our algorithm works for a broader range of parameters, and has a better distance-rate tradeoff for all decoding distances above some moderate thresholds, than the prior algorithms for~$\ell_1$ and~$\ell_2$~\cite{DBLP:journals/tit/RothS94,wu03:_lee_bch_reed_solomon,DBLP:journals/tit/MookP22}; see below for elaboration and \cref{fig:plots} for a visual depiction.
For ease of comparison across the various works and $\ell_{p}$ (semi)metrics, we use a suitably normalized version of distance: for code length~$n$, distance~$d$ corresponds to \emph{relative distance} $\delta \defn d/n^{1/p}$.

For $p=2$, our algorithm can handle an \emph{arbitrarily large} decoding distance, for a correspondingly small enough rate: specifically, as~$\delta$ and the alphabet size grow, we can decode for rates rapidly approaching $1/(\delta\sqrt{2\pi e})$.
By contrast, the prior work~\cite{DBLP:journals/tit/MookP22} applies only for relative distance $\delta < 1/\sqrt{2} \approx 0.7071$ (i.e., $\ell_{2}$ distance less than $\sqrt{n/2}$).
In addition, our algorithm works for larger rates than the one in~\cite{DBLP:journals/tit/MookP22} whenever~$\delta$ exceeds about $0.51797$.
(See \cref{sec:worst-case-ell2} for a detailed comparison.)
This is particularly interesting since the rates obtained in~\cite{DBLP:journals/tit/MookP22} were shown to be \emph{optimal} (in a certain sense) for $\delta < 1/2$, but not for larger values.

For $p=1$, again our algorithm (like the one from~\cite{wu03:_lee_bch_reed_solomon}) can handle an arbitrarily large decoding distance, whereas~\cite{DBLP:journals/tit/RothS94} is limited to relative distance $\delta < 1$ (i.e., $\ell_{1}$ distance less than~$n$).
In addition, our algorithm works for larger rates than those of~\cite{DBLP:journals/tit/RothS94,wu03:_lee_bch_reed_solomon} whenever the relative decoding distance exceeds about $0.78988$, and in general, as~$\delta$ and the alphabet size grow, we can decode for rates rapidly approaching $1/(2e\delta)$.
(See \cref{sec:worst-case-ell1} for details.)
Our algorithm is also qualitatively broader: it decodes from \emph{continuous} (real-valued) error, whereas the ones from~\cite{DBLP:journals/tit/RothS94,wu03:_lee_bch_reed_solomon} require \emph{discrete} (integer) error.
While continuous error can be discretized by rounding, this can increase the relative distance from the codeword by up to $1/2$ in~$\ell_{1}$, which significantly degrades the distance-rate tradeoffs of the prior works, making them worse than ours for all distances.

We also give several useful supplementary results.
By adapting an argument of~\cite{DBLP:journals/tit/RothS94}, we prove lower bounds on the~$\ell_{1}$ and~$\ell_{2}$ minimum distances for a certain natural subclass of GRS codes.
These imply that for many parameters of interest, our list-decoding algorithm outputs at most one codeword, i.e., it is actually a \emph{unique} decoder.
(See \cref{lem:ell2-min-dist,lem:ell1-min-dist} and the discussions thereafter.)
And in addition to \emph{worst-case} errors added by an adversarial channel, we also consider our algorithm's performance under \emph{average-case} errors produced by ``memoryless additive'' channels.
Such channels add independent identically distributed error, drawn from some specified distribution, to each coordinate of the transmitted codeword.
For Laplacian and Gaussian errors (which roughly correspond to~$\ell_{1}$ and~$\ell_{2}$, respectively), we show that our algorithm supports even larger rates than what we would get by merely applying concentration bounds on the error vector and invoking our worst-case results.

\subsection{Technical Overview}%
\label{sec:technical-overview}

At the highest level, our algorithm for list decoding prime-field GRS codes in~$\ell_{p}$ follows the basic approach of~\cite{DBLP:journals/tit/MookP22} for list decoding (G)RS codes in~$\ell_{2}$: we first translate the received word into a suitable \emph{weight} (or \emph{reliability}) vector, then invoke a \emph{soft-decision} list-decoding algorithm~\cite{DBLP:journals/tit/GuruswamiS99,DBLP:phd/us/Guruswami2001,DBLP:journals/tit/KoetterV03a} for GRS codes.
Informally, a weight vector specifies, for each coordinate of the received word and each symbol in the code alphabet, a ``confidence level'' that the transmitted codeword had that symbol at that coordinate.
Given such a weight vector, a soft-decision decoding algorithm then finds all codewords that are sufficiently \emph{correlated} with it, as determined by the code rate.
(For the formal definitions and theorem statement, see \cref{def:weight-vector,def:correlation,thm:GS-soft}.)

For our purposes, the principal challenge is in mapping a received word to an appropriate weight vector so that any codeword that is close enough to the received word (in the~$\ell_{p}$ metric) has sufficient correlation with the weight vector.
The prior work~\cite{DBLP:journals/tit/MookP22} uses very simple weights: given a real received value~$r$, only its floor~$\floor{r}$ and ceiling~$\ceil{r}$ receive positive weights, of $1-(r-\floor{r})$ and $1-(\ceil{r}-r)$, respectively.
It was shown that with these weights, the cited soft-decision algorithms decode up to any~$\ell_{2}$ relative distance $\delta < 1/\sqrt{2}$ for any code rate up to $1-2\delta^{2}$.
Moreover, for $\delta \leq 1/2$ it was shown that this rate is \emph{optimal} for those algorithms, i.e., no other weight assignment can work for a larger rate.
However,~\cite{DBLP:journals/tit/MookP22} did not consider larger decoding distances than these, nor other metrics.

In this work, to handle large decoding distances and other metrics, we use ``smoother'' weights, which typically assign a positive weight to \emph{every} alphabet symbol.
Our overall approach (see \cref{sec:framework}) is quite general, and is parameterized by a function $f \colon \R \to [0,1]$ satisfying mild hypotheses, primarily that its Fourier transform~$\widehat{f}$ is non-negative (see \cref{ass:f-properties}).
This function can be seen as defining a weight---or a relative ``likelihood,'' in the case of a random channel---for every potential real-valued error.\footnote{For example, we take~$f$ to be a Gaussian function for decoding in the~$\ell_{2}$ metric, or under a Gaussian channel.}
For a prime-order code alphabet $\F_{q} \cong \Z_{q} \defn \Z/q\Z$, and a received real value $r \in \R/q\Z$, we assign weight $f(r - w + q\Z) = \sum_{e = r-w \pmod*{q}} f(e)$ to each alphabet symbol $w \in \Z_{q}$.
Since the elements of the coset $r - w + q\Z$ are exactly those errors that would convert a transmitted symbol~$w$ to the received value~$r$, this assignment captures the total weight of all such errors.

Our first main result, given in \cref{thm:corr-lower-bound}, lower bounds the correlation between the weight vector for a received word~$\vecr$ over $\R/q\Z$ and any (code)word~$\vecc$ over~$\Z_{q}$, by the ratio of two quantities determined by~$f$: its (arithmetic or geometric) mean over the coordinates of the error vector $\vecr-\vecc$, and (the square root of) its sum over a certain two-dimensional integer lattice~$\lat_{q}$.
So, for a particular decoding distance in a metric of interest (or channel distribution, in the average case), the goal becomes to choose a suitable function~$f$ that nearly maximizes this ratio.
The proof of the theorem uses a mild generalization of Fourier-analytic results on lattices from~\cite{banaszczyk93:_new,DBLP:journals/siamcomp/MicciancioR07}, and is the source of our requirement that~$\widehat{f}$ is non-negative, and ultimately the restriction that $0 < p \leq 2$ for~$\ell_{p}$ (semi)metrics.

The bulk of the remaining work is then devoted to making a suitable choice of function~$f$ for the~$\ell_{p}$ (semi)metric (and corresponding channel distributions), and analyzing its summation over~$\lat_{q}$.
In \cref{sec:ell_p-norm} we consider scalings of the function $f^{(p)}(x) \defn \exp(-\abs{x}^{p})$, which is known to have non-negative Fourier coefficients for $0 < p \leq 2$ (but not for any other~$p$).
Then in \cref{sec:ell2,sec:ell1} we specialize to $p=2$ and $p=1$, respectively, and give fairly tight upper bounds on $f(\lat_{q})$ using Fourier-analytic techniques or direct analysis.
Finally, we use these bounds to optimize the distance-rate tradeoffs for which we can list decode GRS codes in these~$\ell_{p}$ metrics, and for Gaussian and Laplacian random channels as well.

\section{Preliminaries}%
\label{sec:prelims}

For a positive integer~$n$, let $[n] \defn \set{1,\ldots,n}$.
For a positive integer~$q$, define the quotient ring $\Z_{q} \defn \Z/q\Z$ and the additive quotient group $\R_q \defn \R/q\Z$.
For a prime power~$q$, let $\F_q$ denote the finite field of size~$q$.
When~$q$ is prime, we identify $\Z_{q}$ with the finite field $\F_{q}$ in the natural way.

For any $x \in \R_q$ (which is a coset of $q\Z$), define its ``lift'' $\overline{x} \in [-q/2,q/2)$ to be the unique real number such that $\overline{x} = x \pmod*{q}$, i.e., the ``zero-centered'' distinguished representative of~$x$.
We also apply this notation entry-wise to vectors over~$\R_q$.

For any $p > 0$, define the~$\ell_{p}$ (quasi)norm on $\R^{n}$ as $\norm{\vecx}_{p} \defn \parens[\big]{\sum_{i=1}^{n} \abs{x_{i}}^{p}}^{1/p}$, and for $p=\infty$ define $\norm{\vecx}_{\infty} \defn \max_{i} \abs{x_{i}}$.
It is well known that this is a norm if and only if $p \geq 1$, and it is a \emph{quasinorm} for any $p > 0$.\footnote{A quasinorm relaxes the triangle inequality axiom to require only that $\norm{\vecx+\vecy} \leq K (\norm{\vecx} + \norm{\vecy})$ for some fixed $K$.
  We do not use the triangle inequality, or even this relaxation, so we can consider $p < 1$.}
Similarly, we define the~$\ell_{p}$ (semi)metric on $\R_q^n$ by lifting, i.e., via $\norm{\vecx}_{p} \defn \norm{\overline{\vecx}}_{p}$.\footnote{Formally, this is not a \emph{norm} because it is not defined on a vector space (since $\R_{q}$ is not a field), and it does not satisfy homogeneity due to the mod-$q$ reduction.
  However, it does define a \emph{(semi)metric} (where ``semi'' does not require the triangle inequality), with distance function $d(\vecx,\vecy) = \norm{\vecx-\vecy}_{p}$.}
For $p=1$, this generalizes the Lee metric over $\Z_q$ to $\R_{q}$.

For two groups $X,Y$, their \emph{direct sum} group $X \oplus Y$ is their Cartesian product with the group operation defined component-wise.
This notation extends to the direct sum of group \emph{cosets}, which is a coset of the direct sum of the groups.

For any two vectors $\vecx = (x_1,\ldots,x_n)$ and $\vecy = (y_1,\ldots,y_n)$ of the same dimension, their coordinate-wise (or \emph{Hadamard}) product is denoted by $\vecx \odot \vecy \defn (x_1 \cdot y_1, \ldots, x_n \cdot y_n)$.

For a finite sequence $X_{1}, \ldots, X_{n}$ of real values, we denote their average by $\Avg_{i}[X_{i}] \defn \tfrac{1}{n} \sum_{i=1}^{n} X_{i}$.
We use the following special case of the well known Hoeffding (lower-)tail bound.

\begin{lemma}[Hoeffding's Inequality]%
  \label{lem:Hoeffding}
  Let $X_{1}, \ldots, X_{n}$ be independent identically distributed random variables in $[0,1]$ with common expectation $\mu = \E[X_{i}]$.
  Then for any $\gamma \geq 0$,
  \[ \Pr \bracks[\Big]{\Avg_{i}[X_{i}] \leq \mu - \gamma} < \exp(-2 \gamma^{2} n) \; \text.
  \]
\end{lemma}

\paragraph*{Operations on functions.}

For any function $f \colon D \to \C$ and countable subset $X \subseteq D$, we define $f(X) \defn \sum_{\vecx\in X} f(\vecx)$.
We extend the domain to $D^{k}$ multiplicatively, as
\begin{equation}
  \label{eq:f-multiplicative}
  f^{k}(\vecx) \defn \prod_{i=1}^{k} f(x_i) \; \text,
\end{equation}
often omitting the superscript~$k$ when it is clear from context.
When $D=\R^{n}$, for any real $s \neq 0$ we define $f_{s}(\vecx) \defn f(\vecx/s)$.

\subsection{Linear Codes}%
\label{sec:code-prelims}

A \emph{linear (error-correcting) code} of \emph{(block) length}~$n$ over the \emph{alphabet} $\F_{q}$ is a linear subspace of~$\F_{q}^{n}$.
As a subspace, it has a \emph{dimension}.
In this paper, we consider the following family of codes.

\begin{definition}[(Generalized) Reed--Solomon code]%
  \label{def:GRScode}
  Let $n \leq q$ be positive integers, with~$q$ a prime power.
  For a non-negative integer~$k$, a vector $\vecalpha \in \F_q^n$ with distinct entries, and a vector $\vect \in \parens{\F_q \setminus \set{0}}^n$ with (not necessarily distinct) non-zero entries, the \emph{Generalized Reed--Solomon}~(GRS) code of dimension~$k$ with evaluation points~$\vecalpha$ and twist factors~$\vect$ is defined as
  \[ \grs_{q,k}(\vecalpha,\vect) \defn \set{ \vect \odot f(\vecalpha) = (t_1 \cdot f(\alpha_1) , \ldots, t_n \cdot f(\alpha_n))\ \colon f\in\F_q[x],\ \deg(f)<k} \; \text.\footnote{By convention, the zero polynomial has degree $-\infty$.}
  \]
  A special case is a \emph{Reed--Solomon}~(RS) code, which is obtained by using trivial twist factors $\vect=(1,\ldots,1)$.
\end{definition}

\subsection{Lattices}%
\label{sec:lattice-prelims}

\begin{definition}[Lattice, Basis]%
  \label{def:lat}
  An \emph{($n$-dimensional, full-rank) lattice} $\lat \subset \R^{n}$ is the set of all integer linear combinations of some~$n$ linearly independent \emph{basis} vectors $\matB = \set{\vecb_1,\ldots,\vecb_n} \subset \R^n$:
  \[
    \lat = \lat(\matB) \defn \set[\Big]{\sum_{i=1}^n z_i\vecb_i : z_i \in \Z} \; \text.
  \]
  Equivalently, it is a discrete additive subgroup of $\R^{n}$ whose $\R$-span is $\R^{n}$; as such, it defines the quotient group $\R^{n}/\lat$ of \emph{lattice cosets} $\vecx+\lat$ for $\vecx \in \R^{n}$.
  A sublattice of $\Z^{n}$ is called an \emph{integer lattice}.
\end{definition}

In this work, all lattices are implicitly full rank.
A lattice basis can equivalently be seen as an invertible matrix $\matB \in \R^{n\times n}$ whose columns are the vectors $\vecb_1,\ldots,\vecb_n$.
Note that a given lattice has multiple different bases, which are all related by right-multiplication by \emph{unimodular} matrices in $\Z^{n \times n}$.

\begin{definition}[Determinant]%
  \label{def:det}
  The \emph{determinant} of a lattice~$\lat$ generated by basis~$\matB$ is $\det(\lat) \defn \abs{\det(\matB)}$.
\end{definition}
Note that the determinant of a lattice is invariant under the choice of basis, by the above-mentioned relationship between the bases of a lattice.

\begin{definition}[Dual lattice]%
  \label{def:lat*}
  The \emph{dual lattice} of a lattice $\lat \subset \R^{n}$ is
  \[ \lat^* \defn \set{\vecx\in\R^n : \forall\ \vecv\in\lat,\, \inprod{\vecv}{\vecx}\in\Z} \; \text.
  \] If~$\matB$ is a basis of~$\lat$, then its \emph{dual basis} $\matB^{*} \defn \matB^{-t}$ is a basis of~$\lat^{*}$, and hence $\det(\lat^{*}) = \det(\lat)^{-1}$.
\end{definition}

\begin{lemma}%
  \label{lem:direct-sum-multiplicative}
  Let $f \colon D \to \R$ and $X, Y \subseteq D$ be countable subsets of its domain (e.g., lattice cosets).
  Then $f(X \oplus Y) = f(X) \cdot f(Y)$.
\end{lemma}

\begin{proof}
  This follows directly from the definition of direct sum and multiplicativity (\cref{eq:f-multiplicative}):
  \begin{equation*}
    f(X \oplus Y)
    = \sum_{\substack{\vecx\in X\\ \vecy\in Y}} f(\vecx \oplus \vecy)
    = \sum_{\vecx, \vecy} f(\vecx) \cdot f(\vecy)
    = \Big(\sum_{\vecx} f(\vecx)\Big) \Big(\sum_{\vecy} f(\vecy)\Big)
    = f(X) \cdot f(Y).
  \end{equation*}
\end{proof}

\subsection{Fourier Analysis}%
\label{sec:fourier-prelims}

Let $f \colon \R^n \to \C$ be a (Borel) measurable function that satisfies $\int_{\R^n}\abs{f(\vecx)} \odif{\vecx} <\infty$.
Its \emph{Fourier transform} $\widehat{f} \colon \R^{n} \to \C$ is defined as
\begin{equation*}
  \widehat{f}(\vecw) \defn \int_{\R^n} f(\vecx)\cdot \exp(-2\pi i \inprod{\vecx}{\vecw}) \odif{\vecx} \; \text.
\end{equation*}
It satisfies the following standard properties, which follow by routine calculations.

\begin{lemma}[Multiplicativity]%
  \label{lem:ft-multiplicative}
  For any function~$f$ as above, $\widehat{f^{k}} = \widehat{f}\,^{k}$ (where the exponent notation is as defined in \cref{eq:f-multiplicative}).
\end{lemma}

\begin{lemma}[Time-scaling property]%
  \label{lem:time-scale}
  For any function~$f$ as above and real $s \neq 0$, $\widehat{f_{s}}(\vecw) = s^{n} \cdot \widehat{f}_{1/s}(\vecw)$.
\end{lemma}

\begin{lemma}[{Time-shift property}]%
  \label{lem:time-shift}
  For any function~$f$ as above and $\vecc\in\R^n$, let $g(\vecx) = f(\vecx-\vecc)$.
  Then $\widehat{g}(\vecw) = \widehat{f}(\vecw) \cdot \exp(-2\pi i\inprod{\vecw}{\vecc})$.
\end{lemma}

We say that~$f$ is \emph{nice} if it satisfies conditions that are sufficient for the following formula to hold, e.g., those given in~\cite[pages 106--107]{Serre73}.
All of the specific functions~$f$ we use in this work are easily seen to be nice.

\begin{lemma}[Poisson Summation Formula (PSF)]%
  \label{lem:PSF}
  For any lattice~$\lat$ and nice function~$f$,
  \[ f(\lat) = \det(\lat^*) \cdot \widehat{f}(\lat^*) \; \text. \]
\end{lemma}
We will use a more general version of the PSF for lattice \emph{cosets}.

\begin{lemma}[Generalized PSF]%
  \label{lem:PSF-cosets}
  For any lattice $\lat \subset \R^{n}$, nice function~$f$, and $\vecy \in \R^n$,
  \[ f(\vecy+\lat) = \det(\lat^*)\cdot \sum_{\vecw\in\lat^*} \widehat{f}(\vecw) \cdot \exp(2\pi i\inprod{\vecw}{\vecy}) \; \text.
  \]
\end{lemma}

\begin{proof}
  Define the function $g(\vecx) \defn f(\vecx+\vecy)$.
  By \cref{lem:PSF,lem:time-shift},
  \begin{align*}
    f(\vecy + \lat)
    &= g(\lat) \\
    &= \det(\lat^*) \cdot \widehat{g}(\lat^*) \\
    &= \det(\lat^*) \cdot \sum_{\vecw \in \lat^*} \widehat{g}(\vecw) \\
    &= \det(\lat^*) \cdot \sum_{\vecw \in \lat^*} \widehat{f}(\vecw) \cdot \exp(2\pi i\inprod{\vecw}{\vecy})
      \; \text.
  \end{align*}
\end{proof}

\subsection{Lattice Roughness}%
\label{sec:roughness}

Continuing from \cref{sec:fourier-prelims}, for the rest of this work we require the following properties of~$f$.

\begin{assumption}%
  \label{ass:f-properties}
  The function~$f$ takes values in $[0,1]$ and is nice, and~$\widehat{f}$ is \emph{non-negative real} with $\widehat{f}(0) > 0$.
\end{assumption}

Because~$f$ is real, its Fourier transform is conjugate symmetric, i.e., $\widehat{f}(-w) = \widehat{f}(w)^*$ for all~$w$, where the star denotes complex conjugation.
Since~$\widehat{f}$ is also real, this implies that it is symmetric, i.e., $\widehat{f}(-w) = \widehat{f}(w)$.
Finally, note that if~$f$ satisfies this assumption, then so does its multiplicative extension~$f^{k}$.

We now define a Fourier-analytic quantity that plays an important role in our analysis.
We adopt the name ``roughness'' because it is the functional inverse of the ``smoothing parameter'' from~\cite{DBLP:journals/siamcomp/MicciancioR07}, which is the smallest~$s$ that makes the function $f_s(\vecy+\lat)$ sufficiently ``smooth'' as a function of~$\vecy$.

\begin{definition}%
  \label{def:roughness}
  For a function~$f$, lattice $\lat \subset \R^{n}$, and real $s > 0$, the \emph{roughness} is defined as
  \begin{equation}
    \eps_{\lat,s} \defn \frac{\widehat{f_s}(\lat^* \setminus \set{\veczero})}{\widehat{f_s}(\veczero)}
    = \frac{\widehat{f_s}(\lat^*)}{\widehat{f_s}(\veczero)} - 1 \geq 0.
  \end{equation}
  More generally, for a (linear) subspace~$H$ of $\R^{n}$, the \emph{$H$-roughness} is defined as
  \begin{equation}
    \eps_{\lat,s}(H)
    \defn \frac{\widehat{f_s}(\lat^* \setminus H^\perp)}{\widehat{f_s}(\lat^* \cap H^\perp)}
    = \frac{\widehat{f_{s}}(\lat^{*})}{\widehat{f_{s}}(\lat^{*} \cap H^{\perp})} - 1
    \leq \eps_{\lat,s}(\R^{n}) = \eps_{\lat,s} .
  \end{equation}
  (Both inequalities follow from the non-negativity of $\widehat{f_s}$.)
\end{definition}

\begin{lemma}[{{adapted from~\cite[Lemmas~2.9 and~4.1]{DBLP:journals/siamcomp/MicciancioR07}}}]%
  \label{lem:smoothing}
  For any~$f$ satisfying \cref{ass:f-properties}, lattice $\lat \subset \R^{n}$, real $s>0$, subspace~$H$ of $\R^{n}$ defining roughness $\eps \defn \eps_{\lat,s}(H)$, and $\vecy \in H$, \[ f_s(\vecy + \lat) \in \det(\lat^*) \cdot \widehat{f_s}(\lat^* \cap H^\perp) \cdot [1-\eps, 1+\eps] \; \text, \] with equality against the upper bound when $\vecy=\veczero$.
  In particular, $f_s(\vecy+\lat) \in f_s(\lat)\cdot [\tfrac{1-\eps}{1+\eps},1]$.
\end{lemma}

\begin{proof}
  By the generalized PSF (\cref{lem:PSF-cosets}),
  \begin{align*}
    f_s(\vecy+\lat)
    &= \det(\lat^*) \cdot \sum_{\vecw \in \lat^*} \widehat{f_s}(\vecw) \cdot \exp(2\pi i\inprod{\vecw}{\vecy}) \\
    &= \det(\lat^*) \cdot \parens[\Big]{\widehat{f_s}(\lat^* \cap H^\perp) + \sum_{\vecw \in \lat^* \setminus H^\perp} \widehat{f_s}(\vecw) \cdot \exp(2\pi i \inprod{\vecw}{\vecy})} \\
    &= \det(\lat^*) \cdot \parens[\Big]{\widehat{f_s}(\lat^* \cap H^\perp) + \sum_{\vecw \in \lat^* \setminus H^\perp} \widehat{f_s}(\vecw) \cdot \cos(2\pi \inprod{\vecw}{\vecy})}
      \; \text.
  \end{align*}
  The last equation follows from the symmetry of $\widehat{f_s}$ and by pairing each (non-zero) element of $\lat^* \setminus H^\perp$ with its negation, which cancels out the imaginary part of $\exp(2\pi i \inprod{\vecw}{\vecy})$.

  Now observe that $\widehat{f_s}(\vecw) \cdot \cos(2\pi \inprod{\vecw}{\vecy}) \in [-\widehat{f_s}(\vecw),\widehat{f_s}(\vecw)]$, with equality against the upper bound for $\vecy=\veczero$, because $\widehat{f_s}$ is non-negative.
  The claim then follows by the definition of roughness~$\eps_{\lat,s}(H)$.
\end{proof}

\section{List-Decoding Reed--Solomon Codes}%
\label{sec:framework}

\subsection{Soft-Decision Decoding}%
\label{sec:soft-decision}

To list-decode Reed--Solomon codes under various norms and probabilistic channel models, we use the ``weighted,'' or \emph{soft-decision}, list decoder of Guruswami and Sudan (hereafter GS)~\cite{DBLP:journals/tit/GuruswamiS99}, as elaborated upon in Guruswami's thesis~\cite[Section~6.2.10]{DBLP:phd/us/Guruswami2001} and the work of Koetter and Vardy~\cite{DBLP:journals/tit/KoetterV03a}.
A soft-decision decoder takes as input a vector of non-negative real weights, and outputs a set of codewords.

\begin{definition}%
  \label{def:weight-vector}
  A \emph{weight vector} for a length-$n$ code over~$\F_{q}$ is some $W \defn (W_1,\ldots,W_n) \in \R_{\geq 0}^{qn}$ where each block $W_i \in \R_{\geq 0}^q$ is indexed by~$\F_q$; equivalently, each block is a function $W_i \colon \F_q \to \R_{\geq 0}$.
\end{definition}

Conceptually, each (normalized) block~$W_i$ of a weight vector may be thought of as specifying a (posterior) probability distribution~$\Pi_i$ over $\F_q$, where $\Pi_i(x)$ is the probability that the $i$th transmitted symbol was $x \in \F_q$, given what was received from the channel (which need not be an element of $\F_{q}$).
At a formal level, this interpretation makes sense only when the channel is \emph{probabilistic} (for average-case decoding), but it still serves as useful intuition when the channel is \emph{adversarial} (for worst-case decoding).
We consider both types of channels in our results below.

For $c \in \F_q$, define $[c] \in \bit^q$ to be the binary indicator vector indexed by~$\F_q$ that has a~$1$ in coordinate~$c$ and $0$s elsewhere.
Similarly, for any vector $\vecc=(c_1,\ldots,c_n) \in \F_q^n$, define the weight vector $[\vecc] \defn ([c_1],\ldots,[c_n]) \in \bit^{qn}$.
Observe that its Euclidean norm is $\norm{[\vecc]} = \sqrt{n}$.

\begin{definition}%
  \label{def:correlation}
  The \emph{correlation} between a weight vector $W \in \R_{\geq 0}^{qn}$ and a word $\vecc \in \F_{q}^{n}$ is defined as the length-normalized inner product between~$W$ and~$[\vecc]$ (i.e., the cosine of the angle between them):
  \begin{equation}
    \label{eq:correlation}
    \corr(W,\vecc) \defn
    \frac{\inprod{W}{[\vecc]}}{\norm{W} \cdot \sqrt{n}}
    \; \text.
  \end{equation}
\end{definition}

\begin{theorem}[{{adapted from~\cite[Theorem~18]{DBLP:journals/tit/GuruswamiS99} and~\cite[Theorem~6.21]{DBLP:phd/us/Guruswami2001}}}]%
  \label{thm:GS-soft}
  For a prime power $q$, let $\code \subseteq \F_q^n$ be a Generalized Reed--Solomon code of dimension~$k$ and adjusted rate $R^* \defn (k-1)/n$.
  There is a deterministic algorithm that, given a weight vector~$W$ and a ``tolerance'' $\tau > 0$, outputs in time $\poly(n,q,1/\tau)$ the set of all codewords $\vecc \in \code$ that satisfy
  \begin{equation}
    \label{eq:GS-soft}
    \corr(W,\vecc)
    \geq \sqrt{R^*} + \tau
    \; \text.
  \end{equation}
\end{theorem}

The above follows by taking (in~\cite[Theorem~6.21]{DBLP:phd/us/Guruswami2001}) $\varepsilon = \tau \sqrt{n} \cdot \norm{W} / \norm{W}_{\infty} \geq \tau \sqrt{n}$, so $1/\varepsilon \leq 1/\tau$.
We remark that the theorem was originally stated for \emph{rational} weights, but the supporting argument (from~\cite[Lemma~6.20]{DBLP:phd/us/Guruswami2001}) easily adapts to handle \emph{real}-valued weights that can be lower bounded to any needed precision in polynomial time, as all of ours can be.

\subsection{From Received Words to Weight Vectors}%
\label{sec:received-to-weight}

Here we describe a general approach for translating a received word to a weight vector.
This translation is parameterized by a function that, conceptually, can be viewed as (proportional to) the channel's probability density function, even if the channel is not actually probabilistic.

Let $f \colon \R \to [0,1]$ be a function that satisfies \cref{ass:f-properties}, extended multiplicatively to~$\R^{n}$ as in \cref{eq:f-multiplicative}, and recall that $f_s(x) \defn f(x/s)$ for any constant $s>0$.
Next let~$q$ be a positive integer, and recall that we identify $\Z_q \defn \Z/q\Z$ with~$\F_q$ in the natural way when~$q$ is prime.
Let the set of possible received values be $\R_q = \R/q\Z$, and for any such value $y \in \R_q$, define the weight function $W_{s,y} \colon \Z_{q} \to \R_{\geq 0}$ by
\begin{equation}
  \label{eq:W_y}
  W_{s,y}(x) \defn f_s(y-x + q\Z) \geq 0 \; \text.
\end{equation}
Notice that here~$f_s$ is applied to a \emph{coset} of~$q\Z$, which represents an infinite series; for all our concrete choices, these series converge and so the function $W_{s,y}$ is well defined.
This function can also be seen as the vector $W_{s,y} \defn \parens{W_{s,y}(x)}_{x \in \Z_q} \in \R_{\geq 0}^q$, indexed by~$\Z_q$.

In line with the probabilistic conception of weight vectors from \cref{sec:soft-decision} above, the function $W_{s,y}$ can be seen as follows.
Suppose that a uniformly random symbol in~$\Z_q$ is sent over a channel, which adds (modulo~$q$) noise drawn from a distribution over~$\R$ whose probability density function is proportional to~$f_s$.
Then the probability that the sent symbol was $x \in \Z_q$, conditioned on receiving~$y$, is proportional to $W_{s,y}(x)$.
This is because the coset $y-x \in \R_q$ is the set of all noise values that yield~$y$ if~$x$ is sent.
Note that in the definition of $W_{s,y}$ we do \emph{not} normalize by the total weight $W_{s,y}(\Z_{q}) = f_{s}(y+\Z)$ (which may vary based on the received value~$y$); this turns out to yield simpler analyses and tighter results.

\begin{definition}%
  \label{def:received-weight}
  For a function~$f_s$ as above and any received vector $\vecy = (y_1, \ldots, y_n) \in \R_q^n$, define the corresponding weight vector as
  \begin{equation}
    \label{eq:weight}
    W_{s,\vecy} \defn (W_{s,y_1}, \ldots, W_{s,y_n}) \in \R_{\geq 0}^{nq}
    \; \text.
  \end{equation}
\end{definition}

In order to use the soft-decision algorithm (\cref{thm:GS-soft}) for decoding under an adversarial channel, it suffices to show that we can choose a suitable~$s$ so that for any received word~$\vecy$ and any sufficiently close codeword~$\vecc$ (in the norm of interest), the correlation $\corr(W_{s,\vecy},\vecc)$ satisfies~\eqref{eq:GS-soft}.
Similarly, for decoding under a probabilistic channel, it suffices to show that with high probability over the channel noise~$\vece$, the transmitted codeword~$\vecc$ has large enough correlation with the weight vector~$W_{s,\vecy}$ of the received word $\vecy = \vecc + \vece$ (again, for some suitably chosen~$s$).
To this end, in what follows we give a lower bound on $\inprod{W_{s,\vecy}}{[\vecc]}$ and an upper bound on $\norm{W_{s,\vecy}}$, in terms of~$f_s$ and the difference $\vecy-\vecc$ between the received word and the codeword of interest.

\subsection{Main Theorem}%
\label{sec:main-thm}

Here we state and prove the main result of this section.
For this we define the two-dimensional integer lattice~$\lat_{q}$ that consists of all shifts of the lattice $q\Z^2$ by $(z,z)$ for an integer~$z$, i.e.,
\begin{equation}
  \lat_{q} \defn \bigcup_{x \in \Z_q} (x \oplus x) = \bigcup_{z \in \Z} ((z,z) + q\Z^2) \supset q\Z^2 .
\end{equation}
We have that $\det(\lat_{q})=q$, and so $\det(\lat_{q}^*)=1/q$.
We sometimes omit the~$q$ subscript when it is clear from context or its value is unimportant.

\begin{theorem}%
  \label{thm:corr-lower-bound}
  Suppose that~$f$ satisfies \cref{ass:f-properties}.
  For any $s>0$ and $\vecy \in \R_q^n$ defining $W=W_{s,\vecy}$, and any $\vecc \in \Z_q^n$,
  \[ \corr(W,\vecc)
    \geq \frac{\Avg_{i \in [n]}[f_{s}(y_{i}-c_{i})]}{\sqrt{f_{s}(\lat_{q})}}
    \geq \frac{f_s(\vecy-\vecc)^{1/n}}{\sqrt{f_s(\lat_{q})}}
    \; \text. \]
\end{theorem}

\begin{proof}
  This follows immediately from the following lower and upper bounds on the numerator and denominator of $\corr(W,\vecc) = \frac{\inprod{W}{[\vecc]}/n}{\norm{W}/\sqrt{n}}$.
  For the numerator, by the definitions of~$W$ and~$[\vecc]$,
  \begin{equation*}
    \inprod{W}{[\vecc]}/n
    = \Avg_{i \in [n]}\bracks*{W_{s,y_i}(c_i)}
    = \Avg_{i \in [n]}\bracks*{f_s(y_i-c_i)}
    \geq f_{s}\parens{\vecy-\vecc}^{1/n}
    \; \text,
  \end{equation*}
  where the last step follows by the inequality of arithmetic and geometric means, and the non-negativity and multiplicativity of~$f_{s}$ over direct sums of cosets (\cref{lem:direct-sum-multiplicative}).
  For the denominator, the upper bound $\norm{W}/\sqrt{n} \leq \sqrt{f_{s}(\lat_{q})}$ is proved in \cref{lem:W.W} below.
\end{proof}

\begin{remark}%
  \label{rem:restriction}
  If certain coordinates of~$\vecy$ are ``very far'' from the corresponding entries of~$\vecc$, we may get a better lower bound on $\corr(W,\vecc)$ by \emph{restricting} to ``good'' coordinates.
  Specifically, for any nonempty $G \subseteq [n]$ of cardinality $g=\abs{G}$, by the non-negativity of~$f_{s}$, \[ \Avg_{i \in [n]}[f_{s}(y_{i}-c_{i})] \geq \frac{g}{n} \cdot \Avg_{i \in G}[f_{s}(y_{i}-c_{i})] \geq \frac{g}{n} \cdot f_{s}\parens{\vecy_{G}-\vecc_{G}}^{1/g} \; \text,
  \]
  where~$\vecx_{G}$ is the vector obtained by restricting~$\vecx$ to the coordinates in~$G$.
\end{remark}

\begin{lemma}%
  \label{lem:W.W}
  Adopting the setup from \cref{thm:corr-lower-bound}, and letting $\tilde{\eps} = \eps_{\lat_{q},s}(H)$ where $H=\spn(1,1)$,
  \begin{equation*}
    \norm{W}^{2}/n
    \in f_s(\lat_{q}) \cdot \bracks[\Big]{\frac{1-\tilde{\eps}}{1+\tilde{\eps}}, 1}
    \; \text.
  \end{equation*}
\end{lemma}

\begin{proof}
  By definition of~$W$,
  \begin{equation*}
    \norm{W}^{2}/n
    = \Avg_{i \in [n]}\bracks[\Big]{\sum_{x\in\Z_q} f_s\parens{y_i-x}^2}
    \; \text.
  \end{equation*}
  To bound this, let $y\in\R_q$ be arbitrary.
  By \cref{lem:direct-sum-multiplicative},
  \begin{align*}
    \sum_{x \in \Z_q} f_s\parens{y-x}^2
    &= \sum_{x \in \Z_q} f_s((y-x) \oplus (y-x))
    \\ &= \sum_{x \in \Z_q} f_s((y \oplus y) - (x \oplus x))
    \\ &= f_{s}((\overline{y},\overline{y}) + \lat_{q})
    \\ &\in f_s(\lat_{q}) \cdot \bracks[\Big]{\frac{1-\tilde{\eps}}{1+\tilde{\eps}}, 1}
      \; \text,
  \end{align*}
  where the last step follows by the latter part of \cref{lem:smoothing} on the lattice~$\lat_{q}$ with subspace~$H$, and noting that $(\overline{y},\overline{y}) \in H$.
  The claim follows by averaging over $i \in [n]$.
\end{proof}

\subsection{Average-Case Decoding}%
\label{sec:average-case-decod}

Here we consider list-decoding in the \emph{average case}, where the channel is probabilistic (not worst case) and the goal is to output a list of codewords that includes the transmitted one.
We consider channels that add independent, identically distributed random error (drawn from some specified distribution) to each coordinate of the transmitted codeword; this is often known as a \emph{memoryless additive channel}.
Specifically, we assume that the channel's error distribution (for each coordinate) is proportional to~$f_{r}$ for some $r > 0$, i.e., it has probability density function
\begin{equation}
  \label{eq:D_r}
   D_{r}(x) \defn \frac{f_{r}(x)}{\widehat{f_{r}}(0)} \; \text.
\end{equation}
For example, if~$f_{r}$ is a Gaussian function, this is known as the \emph{additive white Gaussian noise} (AWGN) channel model.
In some settings one may also consider a \emph{discrete} channel distribution, e.g., over~$\Z$, in which case its probability mass function is $D_{r}(x) \defn f_{r}(x)/f_{r}(\Z)$.
For any $s > 0$ (which may differ from~$r$), define
\begin{equation}
  \label{eq:mu_rs}
  \mu_{r,s} \defn \E_{e \gets D_{r}}[f_{s}(e)] \; \text.
\end{equation}

In \cref{sec:ell_p-norm} we will use the following bound for a specific family of functions~$f$ to show that the transmitted codeword is recovered with high probability over the channel error.

\begin{lemma}%
  \label{lem:avg-case-generic}
  For any $r, s > 0$ and~$T$ defining $\gamma \defn \mu_{r,s} - T \cdot \sqrt{f_{s}(\lat_{q})} \geq 0$, and any $\vecc \in \Z_q^n$,
  \[ \Pr_{\vece \gets D_{r}^{n}} \bracks[\big]{\corr(W_{s,\vecc+\vece},\vecc) \leq T} < \exp(-2 \gamma^{2} n) \; \text. \]
\end{lemma}

\begin{proof}
  The error coordinates~$e_{i}$ are drawn independently from $D_{r}$, so by \cref{ass:f-properties} the values $f_{s}(e_{i}) \leq f_{s}(e_{i} + q\Z)$ are independent and identically distributed random variables in $[0,1]$, with expected value~$\mu_{r,s}$.
  Then by \cref{thm:corr-lower-bound} and Hoeffding's inequality (\cref{lem:Hoeffding}),
  \begin{align*}
    \Pr_{\vece} \bracks{\corr(W_{s,\vecc+\vece},\vecc) \leq T}
    &\leq \Pr_{\vece} \bracks[\Bigg]{\Avg_{i \in [n]}[f_s(e_i)] \leq T \cdot \sqrt{f_{s}(\lat_{q})}}
    \\ &< \exp(-2 \gamma^{2} n)
         \; \text.
  \end{align*}
\end{proof}

\section{General $\ell_{p}$ (Semi)Metrics}%
\label{sec:ell_p-norm}

In this section we define weight vectors via \cref{def:received-weight} using the function $f \colon \R \to [0,1]$ defined as
\begin{align}
  \label{eq:fp}
  f(x) = f^{(p)}(x) &\defn \exp\parens[\big]{- \parens{c_{p} \abs{x}}^{p}} \\
  \text{where } c_{p} &\defn 2\cdot \Gamma(1 + 1/p) \notag
  \; \text,
\end{align}
where the gamma function $\Gamma(z) = \int_{0}^{\infty} u^{z-1} \exp(-u) \odif{u}$ for $z > 0$, and satisfies $\Gamma(1)=1$ and $\Gamma(1+z)=z \cdot \Gamma(z)$.
As two important examples, $c_{1} = 2$ and $c_{2} = \sqrt{\pi}$.

Note that by multiplicativity (\cref{eq:f-multiplicative}), \[ f(\vecx) = \prod_{i=1}^{n} f(x_{i}) = \exp\parens[\Big]{-\sum_{i=1}^{n} \parens{c_{p} \abs{x_{i}}}^{p}} = \exp\parens[\big]{-\parens{c_{p} \length{\vecx}_{p}}^{p}} = f(\norm{\vecx}_{p}) \; \text.
\]
Regarding the Fourier transform of~$f$, the ``normalizing constant''~$c_{p}$ has been defined to make $\widehat{f}(0) = 1$:
\[ \widehat{f}(0)
  = \int_{-\infty}^{\infty} f(x) \odif{x}
  = 2 \int_{0}^{\infty} \exp \parens[\big]{-\parens{c_{p} x}^{p}} \odif{x}
  = \frac{2}{p \cdot c_{p}} \int_{0}^{\infty} u^{1/p - 1} \exp(-u) \odif{u} = \frac{2 \cdot \Gamma(1/p)}{p \cdot c_{p}}
  = 1 \; \text,
\]
by the change of variable $u = \parens{c_{p} x}^{p}$.

It is also known that~$\widehat{f}$ is non-negative for $0 < p \leq 2$; this follows immediately from an elegant lemma and proof due to Logan, given in~\cite[Lemma~5]{elkies91:_packing_densities}.\footnote{For $p > 2$, by contrast,~$\widehat{f}$ can have negative values, which prevents our framework from supporting~$\ell_p$ metrics for such~$p$.}
So,~$f$ satisfies \cref{ass:f-properties} for such~$p$.
Another immediate consequence of Logan's lemma is that as~$s$ grows, $\widehat{f_{s}}(w)/s$ strictly decreases and approaches zero for every $w \neq 0$.\footnote{Let $0 < p \leq 2$ and $h(x) = \exp(-\abs{x}^{p})$; since $f(x) = h(c_p x)$, it suffices to prove the claimed property for~$h_s$.
  Logan shows that $h_{s}(x)$ can be expressed as a non-negative linear combination of Gaussians, as $h_{s}(x) = \int_0^\infty \exp(-t \parens{x/s}^2) \odif \alpha(t)$, where $\alpha(t)$ is bounded and non-decreasing.
  For every $t>0$, the Fourier transform (with respect to $x$) of $\exp(-t \parens{x/s}^2)/s$ is $\exp(-\parens{\pi s w}^{2}/t) \cdot \sqrt{\pi/t}$.
  As~$s$ increases, this strictly decreases and approaches zero for any $w \neq 0$.
  Since $\alpha(t)$ is bounded and non-decreasing, the same goes for  $\widehat{h_s}(w)/s$.}

\subsection{Worst-Case Decoding}%
\label{sec:worst-ell_p}

We now address list-decoding in the~$\ell_{p}$ (semi)metric for $0 < p \leq 2$, under worst-case error.
Consider decoding distance $d = \delta \cdot n^{1/p}$, where~$n$ is the code length, and~$\delta$ can be seen as the \emph{relative} decoding distance (relative to $n^{1/p}$, which is the most natural normalization factor for~$\ell_{p}$).
For $s > 0$, relative distance $\delta \geq 0$, and positive integer modulus~$q$, define
\begin{equation}
  \label{eq:Bpqdelta}
  W^{(p)}_{q,\delta}(s)
  \defn \frac{f_{s}(\delta)}{\sqrt{f_{s}(\lat_{q})}}
  = \frac{\exp\parens[\big]{- \parens{c_{p} \cdot \delta / s}^{p}}}{\sqrt{f_{s}(\lat_{q})}}
  \geq 0
  \; \text.
\end{equation}

By \cref{thm:corr-lower-bound,thm:GS-soft}, to decode a GRS code of adjusted rate~$R^{*}$ over a prime field~$\F_{q}$ to within $\ell_{p}$ distance $\delta \cdot n^{1/p}$ using the GS algorithm, it suffices to set $s > 0$ so that $W^{(p)}_{q,\delta}(s) > \sqrt{R^{*}}$.
In other words, we can decode under relative distance~$\delta$ for any~$R^{*}$ less than
\begin{equation}
  \label{eq:rate-versus-distance-ell_p}
  R^{*,(p)}_{\text{wc},q}(\delta)
  \defn \sup_{s > 0} W^{(p)}_{q,\delta}\parens{s}^{2}
  \; \text.
\end{equation}
The following makes this formal.

\begin{theorem}%
  \label{thm:worst-ell_p}
  For any $0 < p \leq 2$, $\delta \geq 0$, and prime~$q$, the GS soft-decision algorithm using weight vector given by $f^{(p)}_{s}$ for any $s > 0$ list-decodes, up to~$\ell_p$ distance $d = \delta \cdot n^{1/p}$, any GRS code $\code\subseteq \F_q^n$ with adjusted rate $R^* < W^{(p)}_{q,\delta}\parens{s}^{2}$, in time polynomial in~$n$, $q$, and $1/(W^{(p)}_{q,\delta}(s) - \sqrt{R^{*}})$.\footnote{We remark that in many cases, the concrete polynomial running time can be improved using a lower bound for $\norm{W}_{2}$, such as the one given by \cref{lem:W.W}.}
\end{theorem}

\begin{proof}
  We invoke the GS algorithm on the weight vector $W = W_{s,\vecy}$ given by the choice of~$s$ and the received word~$\vecy$, and tolerance $\tau = W^{(p)}_{q,\delta}\parens{s} - \sqrt{R^{*}} > 0$.\footnote{To be more precise, we can invoke GS on any approximation of~$\tau$ in $[\tau/2,\tau]$, say.
    This can be computed by approximating $f_{s}(\lat_{q})$ from above to the needed precision, by enumerating sufficiently many points of~$\lat_{q}$ near the origin, and upper-bounding the contribution of the remaining points in the ``tails'' using, e.g., \cref{lem:Banaszczyk}.}
  The running time is polynomial in~$n$, $q$, and $1/\tau$.

  Now let $\vecc \in \code$ be a codeword within distance~$d$ of~$\vecy$, i.e., $\norm{\overline{\vecy-\vecc}}_{p} \leq d$.
  By \cref{thm:corr-lower-bound,ass:f-properties,eq:Bpqdelta},
  \begin{equation*}
    \corr(W,\vecc)
    \geq \frac{f_{s}\parens{\vecy-\vecc}^{1/n}}{\sqrt{f_s(\lat_{q})}}
    \geq \frac{f_{s}\parens{d}^{1/n}}{\sqrt{f_{s}(\lat_{q})}}
    = W^{(p)}_{q,\delta}(s) = \sqrt{R^{*}} + \tau
    \; \text.
  \end{equation*}
  So, by \cref{thm:GS-soft}, the output of the GS algorithm includes~$\vecc$, as needed.
\end{proof}

\begin{remark}%
  \label{rem:restriction-wc-ell_p}
  Following \cref{rem:restriction}, suppose that the received word~$\vecy$ is within relative distance~$\delta$ of a codeword~$\vecc$ on some subset $G \subseteq [n]$ of $g=\abs{G}$ coordinates, i.e., $\norm{\overline{\vecy_{G} - \vecc_{G}}} \leq \delta \cdot g^{1/p}$, and the remaining coordinates of~$\vecy$ may be \emph{arbitrary}.
  Then $\corr(W_{s,\vecy},\vecc) \geq (g/n) \cdot W^{(p)}_{q,\delta}(s)$.
  So, we can list-decode all such codewords as long as the adjusted rate $R^{*} < \parens{g/n}^{2} \cdot R^{*,(p)}_{\text{wc},q}(\delta)$.
\end{remark}

\begin{remark}%
  \label{rem:optimize-ell_p}
  Interestingly, as~$\delta$, $q/\delta$, and~$n$ grow (and the other parameters remain fixed), the product of the relative distance~$\delta$ and the adjusted rate~$R^{*}$ for which we can decode approaches the \emph{relative radius of a unit-volume~$\ell_{p}$ ball}.
  To see this, first observe that as $q/s$ grows, $f_{s}(\lat_{q})$ approaches \[ f_{s}(\lat_{q} \cap \spn(1,1)) = \sum_{z \in \Z} f_{s}(z,z) = \sum_{z \in \Z} f_{s/2^{1/p}}(z) = f_{s/2^{1/p}}(\Z) = \widehat{f_{s/2^{1/p}}}(\Z) \; \text,
  \]
  where the last equality is by the PSF (\cref{lem:PSF}).
  Then as~$s$ grows, the above approaches $s/2^{1/p}$, because $\widehat{f}(0)=1$ and hence $\widehat{f_{s/2^{1/p}}}(0) = s/2^{1/p}$ by \cref{lem:time-shift}, and the other Fourier coefficients approach zero.

  So, as~$s$ and~$q/s$ grow, the bound $W^{(p)}_{q,\delta}\parens{s}^{2}$ on the adjusted rate approaches $2^{1/p} \cdot \exp\parens[\big]{-\parens{2^{1/p} c_{p} \cdot \delta/s}^{p}}/s$.
  A straightforward calculation using the change of variable $t=\parens{2^{1/p} c_{p} \cdot \delta/s}^{p}$ shows that this is maximized for $t=1/p$ (and hence $s=\parens{2p}^{1/p} c_{p} \cdot \delta$), so we can decode to within relative~$\ell_{p}$ distance~$\delta$ for an adjusted rate~$R^{*}$ approaching
  \[ \frac{1}{\parens{ep}^{1/p} \cdot c_{p} \cdot \delta} \; \text. \]

  By comparison, it is known that the volume of an $n$-dimensional~$\ell_{p}$ ball of unit relative radius (i.e., radius $n^{1/p}$) has $n$th root
  \[ 2 n^{1/p} \cdot \frac{\Gamma(1+1/p)\phantom{^{1/n}}}{\Gamma\parens{1+n/p}^{1/n}} \longrightarrow \parens{ep}^{1/p} \cdot c_{p} \; \text, \] using Stirling's approximation for the denominator as~$n$ grows.
  So, the relative radius of a unit-volume~$\ell_{p}$ ball is the reciprocal of this, which is what $R^{*} \cdot \delta$  approaches.
\end{remark}

\subsection{Average-Case Decoding}%
\label{sec:avg-case-ell_p}

We now consider average-case decoding under a memoryless additive (continuous or discrete) channel whose density function is proportional to a scaling of $f = f^{(p)}$.
Specifically, we consider the continuous distribution with probability density function $D_{r}(x) \defn f_{r}(x)/r$, and the discrete distribution over~$\Z$ with probability mass function $D_{r}(x) \defn f_r(x) / f_r(\Z)$.
Following \cref{sec:average-case-decod}, for any $r,s > 0$ define \[ \mu^{(p)}_{r,s} \defn \mu_{r,s} = \E_{e \gets D_{r}}[f_{s}(e)] \; \text.
\]
For these channel distributions we derive suitable bounds on $\mu^{(p)}_{r,s}$, then reach the conclusion via \cref{lem:avg-case-generic,thm:GS-soft}.

\begin{lemma}%
  \label{lem:mu-bound-ell_p}
  For any $0 < p \leq 2$, any $r > 0$ defining a continuous or discrete distribution~$D_r$, and $s > 0$,
  \[ \mu^{(p)}_{r,s} \geq \frac{s}{\norm{(r,s)}_{p}}
    \; \text{,} \]
  with equality in the continuous case and strict inequality in the discrete case.
\end{lemma}

\begin{proof}
  First note that for $t \defn (rs)/\parens{r^{p}+s^{p}}^{1/p}$, we have $f_{r}(x) \cdot f_{s}(x) = f_{t}(x)$, since $1/t^{p}=1/r^{p}+1/s^{p}$.
  For the continuous case,
  \begin{equation*}
    \mu^{(p)}_{r,s}
    = \int_{\R} D_r(e) \cdot f_s(e) \odif{e}
    = \frac{1}{r} \int_{\R} f_{t}(e) \odif{e}
    = \frac{t}{r}
    = \frac{s}{\norm{(r,s)}_{p}} \; \text.
  \end{equation*}
  For the discrete case, since $t < r$, we have that $\widehat{f_{t}}(w)/t > \widehat{f_{r}}(w)/r$ for all $w \in \Z \setminus \set{0}$ (and equality at $w=0$), so by \cref{lem:PSF},
  \begin{equation*}
    \mu^{(p)}_{r,s}
    = \sum_{e \in \Z} D_r(e) \cdot f_s(e)
    = \frac{f_{t}(\Z)}{f_r(\Z)}
    = \frac{t}{r} \cdot \frac{\widehat{f_{t}}(\Z)/t}{\widehat{f_{r}}(\Z)/r}
    > \frac{t}{r}
    \; \text.
  \end{equation*}
\end{proof}

Now, for any channel parameter $r > 0$ and for $s > 0$, define
\begin{equation}
  \label{eq:Apqr}
  A^{(p)}_{q,r}(s) \defn \frac{\mu^{(p)}_{r,s}}{\sqrt{f_{s}(\lat_{q})}}
  \geq \frac{s}{\norm{(r,s)}_{p} \cdot \sqrt{f_{s}(\lat_{q})}}
  \; \text,
\end{equation}
where the inequality is by \cref{lem:mu-bound-ell_p}.
By \cref{thm:corr-lower-bound,thm:GS-soft}, to decode (with high probability) a GRS code of adjusted rate~$R^{*}$ over a prime field $\F_{q}$ under a channel with parameter~$r$, it suffices to set $s > 0$ so that $A^{(p)}_{q,r}(s) > \sqrt{R^{*}}$.
In other words, we can decode under channel parameter~$r$ for any~$R^{*}$ less than
\begin{equation}
  \label{eq:rate-versus-channel-ell_p}
  R^{*,(p)}_{\text{ac},q}(r) \defn \sup_{s > 0} A^{(p)}_{q,r}\parens{s}^{2}
  \; \text.
\end{equation}
The following makes this formal.

\begin{theorem}%
  \label{thm:avg-ell_p}
  Let $0 < p \leq 2$, $r > 0$, $\alpha \in (0,1)$, and~$q$ be prime.
  Under a memoryless additive (continuous or discrete) channel with distribution $D_{r}$, the GS soft-decision algorithm, using weight vector given by $f^{(p)}_{s}$ for any $s > 0$, list-decodes any GRS code $\code \subseteq \F_q^n$ with adjusted rate $R^* < A^{(p)}_{q,r}\parens{s}^{2}$, in time polynomial in~$n$, $q$, and $1/(A^{(p)}_{q,r}(s) - \sqrt{R^{*}})$, except with probability less than
  \[ \exp\parens[\big]{-2n \cdot f_{s}(\lat_{q}) \cdot \alpha^{2} \cdot \parens[\big]{A^{(p)}_{q,r}(s)-\sqrt{R^{*}}}^{2}}
    \; \text.
  \]
\end{theorem}

\begin{proof}
  Throughout the proof let $A(s) \defn A^{(p)}_{q,r}(s)$.
  We invoke the GS algorithm on the weight vector $W = W_{s,\vecy}$ given by the choice of~$s$ and the received word~$\vecy$, and tolerance $\tau = T - \sqrt{R^{*}} > 0$, where
  \begin{align}
    T
    &= A(s) - \alpha(A(s) - \sqrt{R^{*}})
      \label{eq:T1}
    \\ &= \sqrt{R^{*}} + (1-\alpha) (A(s) - \sqrt{R^{*}}) \in (\sqrt{R^{*}}, A(s))
      \; \text.
      \label{eq:T2}
  \end{align}
  The running time is polynomial in~$n$, $q$, and $1/\tau = \Theta(1/(A(s) - \sqrt{R^{*}}))$.
  
  Now suppose that $\vecy = \vecc + \vece$, where $\vecc \in \code$ is the transmitted codeword and $\vece \gets D_{r}^n$ is the channel error.
  To show that the list output by the GS algorithm contains~$\vecc$ with the claimed probability, by \cref{thm:GS-soft} it suffices to show that $\corr(W,\vecc) \geq \sqrt{R^{*}} + \tau = T$.
  Following the setup of \cref{lem:avg-case-generic}, let
  \begin{align*}
    \gamma
    &= \mu_{r,s} - T \cdot \sqrt{f_{s}(\lat_{q})}
    \\ &= \sqrt{f_s(\lat_{q})} \cdot \parens{A(s) - T}
       & \text{(\cref{eq:Apqr})}
    \\ &= \sqrt{f_s(\lat_{q})} \cdot \alpha \parens{A(s) - \sqrt{R^*}} > 0
       & \text{(\cref{eq:T1}).}
  \end{align*}
  So, by \cref{lem:avg-case-generic}, $\Pr_{\vece}[\corr(W,\vecc) < T] < \exp(-2\gamma^2 n)$, which yields the claim.
\end{proof}

\begin{remark}
  \Cref{thm:avg-ell_p} outperforms \cref{thm:worst-ell_p} (for worst-case decoding) by a factor that approaches $\parens{e/2}^{1/p}$ in the adjusted rate~$R^{*}$ it can handle, as~$r$ and $q/r$ grow.
  Specifically, consider a channel with parameter~$r$.
  A calculation reveals that its relative error (in~$\ell_{p}$, relative to~$n^{1/p}$) is tightly concentrated around $\delta = r/(p^{1/p} \cdot c_{p})$, so following the analysis in \cref{rem:optimize-ell_p}, \cref{thm:worst-ell_p} applies for~$R^{*}$ that approaches $1/(r \cdot e^{1/p})$.
  By comparison, \cref{thm:avg-ell_p} applies for~$R^{*}$ that approaches $1/(r \cdot 2^{1/p})$.
\end{remark}

\begin{remark}%
  \label{rem:restriction-ac-ell_p}
  Following \cref{rem:restriction}, and similarly to \cref{rem:restriction-wc-ell_p}, suppose that there exists a subset $G \subseteq [n]$ of $g=\abs{G}$ ``good'' coordinates for which the channel generates the received word~$\vecy$ from the transmitted codeword~$\vecc$ by adding independent noise from distribution~$D_{r}$ on those coordinates, and sets the remaining coordinates \emph{arbitrarily}.
  Then our lower bound on $\corr(W_{s,\vecy},\vecc)$ from \cref{thm:corr-lower-bound} involves an extra $g/n$ factor, and an average over just the coordinates in~$G$.
  So, we can correctly list-decode for any adjusted rate $R^{*} < \parens{g/n}^{2} \cdot A^{(p)}_{q,r}\parens{s}^{2}$.
  More precisely, the statement and proof of \cref{thm:avg-ell_p} hold with every occurrence of $A^{(p)}_{q,r}(s)$ having an additional $g/n$ factor, and with an extra $n/g$ factor inside the $\exp$ expression for the failure probability.
\end{remark}

\section{The $\ell_2$ Metric and Gaussian Error}%
\label{sec:ell2}

In the remainder of the paper we instantiate our general list-decoding results for~$\ell_{p}$ (semi)metrics (\cref{thm:worst-ell_p,thm:avg-ell_p}) to specific metrics of interest and memoryless additive channels.
In this section, we consider the~$\ell_2$ metric and Gaussian channels.

We specialize \cref{eq:fp} to $p=2$, i.e., the Gaussian function
\[ f(x) \defn f^{(2)}(x) = \exp(-\pi x^2) \; \text.
\]
By a straightforward calculation it can be seen that this function is its own Fourier transform: $\widehat{f} = f$.
Note that $\widehat{f_s} = s \cdot f_{1/s}$ by the time-scaling property of the Fourier transform (\cref{lem:time-scale}).
Finally, recalling that $f(\vecx) = f(\norm{\vecx}_{2})$, we get that~$f$ is invariant under rotations.

\subsection{Bounds}%
\label{sec:bounds-ell2}

In this subsection we derive fairly tight bounds on the factor $f_{s}(\lat_{q})$ that appears in the quantities that govern the adjusted rates under which we can decode in the worst and average cases (\cref{eq:Bpqdelta,eq:Apqr}, respectively).
For this purpose we need to define a suitable ``fudge factor.''
For $r \geq r_{0} \defn \sqrt{\ln(4)/\pi} \approx 0.66428$, define
\begin{equation}
  \label{eq:E}
  E(r) \defn 1 - 2\exp(-\pi r^{2}/2) \in [0,1) \; \text.
\end{equation}
Notice that $E(r)$ is positive for $r > r_{0}$, is strictly increasing, and rapidly approaches~$1$ as~$r$ increases.
Next, for real $s, q$ such that $s \in [r_{0}, q/r_{0}]$, define
\begin{equation}
  \label{eq:Ep}
  E_{q}(s) \defn \sqrt{E(q/s) \cdot E(s)} \in [0,1) \; \text.
\end{equation}
Similarly, $E_{q}(s)$ is positive for $s \in (r_{0}, q/r_{0})$, and rapidly approaches~$1$ as both $s, q/s$ increase.

\begin{lemma}%
  \label{lem:common-ell2}
  For any real~$s$ and positive integer~$q$ such that $s \in (r_{0}, q/r_{0})$,
  \[ \frac{1}{f_{s}(\lat_{q})} > \frac{\sqrt{2}}{s} \cdot E_{q}\parens{s}^{2} \; \text. \]
\end{lemma}

\begin{proof}
  This follows directly from \cref{lem:fsL-ell2,lem:roughness-bounds} below.
  Specifically, let $\eps' = \eps_{\Z, q/(s\sqrt{2})}$ and $\tilde{\eps} = \eps_{\lat_{q},s}(H)$.
  By \cref{lem:roughness-bounds} (applied twice, with $r=q/s$ and $r=s$, which are both greater than~$r_{0}$),
  \begin{equation*}
    \frac{1}{(1 + \eps')(1 + \tilde{\eps})} > E_{q}\parens{s}^{2}
    \; \text.
  \end{equation*}
  The result then follows by \cref{lem:fsL-ell2}.
\end{proof}

\begin{lemma}%
  \label{lem:fsL-ell2}
  For any real $s > 0$ and positive integer~$q$, let $\eps' = \eps_{\Z,q/(s\sqrt{2})}$ and $\tilde{\eps} = \eps_{\lat_{q},s}(H)$ where $H=\spn(1,1)$.
  Then
  \[ f_s(\lat_{q}) = \frac{s}{\sqrt{2}} \cdot (1+\eps') \cdot (1+\tilde{\eps}) \; \text.
  \]
\end{lemma}

\begin{proof}
  Recall that $\lat = \lat_{q}$ has determinant $\det(\lat)=q$, hence its dual has determinant $\det(\lat^{*}) = 1/q$.
  A basis for~$\lat$ consists of the vectors $(1,1)$ and $(q,0)$, and its dual basis consists of the vectors $(0,1)$ and $(1,-1)/q$.

  Since $H^{\perp} = \spn(1,-1)$, we have that $\lat^{*} \cap H^{\perp}$ consists merely of all the integer multiples of the dual basis vector $(1,-1)/q$, which has Euclidean norm $\sqrt{2}/q$.
  Therefore, $\lat^{*} \cap H^{\perp}$ is a rotation of $(\sqrt{2}/q) \Z$.
  So,
  \begin{align*}
    f_{s}(\lat)
    &= (1/q) \cdot \widehat{f_{s}^{2}}(\lat^{*} \cap H^{\perp}) \cdot (1+\tilde{\eps})
    & \text{(\cref{lem:smoothing} and definition of $\tilde{\eps}$)}
    \\ &= (s^{2}/q) \cdot f_{1/s}^{2}(\lat^{*} \cap H^{\perp}) \cdot (1+\tilde{\eps})
    & \text{(\cref{lem:time-scale})}
    \\ &= (s^{2}/q) \cdot f_{q/(s\sqrt{2})}(\Z) \cdot (1+\tilde{\eps})
    & \text{(rotational invariance of~$f^{2}$, rescaling)}
    \\ &= (s/\sqrt{2}) \cdot (1+\eps') \cdot (1+\tilde{\eps})
    & \text{(\cref{lem:smoothing} and definition of $\eps'$).}
  \end{align*}
\end{proof}

Next we bound the roughness quantities $\eps', \tilde{\eps}$ from \cref{lem:common-ell2,lem:fsL-ell2}, using the following classic tail inequality.

\begin{lemma}[{{adapted from~\cite[Lemma~2.4]{DBLP:journals/dcg/Banaszczyk95}}}]%
  \label{lem:Banaszczyk}
  For any lattice~$\lat$, unit vector $\vecu$, and $s,t > 0$,
  let $T_{\vecu,t} = \set{\vecx : \abs{\inprod{\vecx}{\vecu}} \geq t}$.
  Then
  \[ f_s(\lat \cap T_{\vecu,t}) < 2\exp(-\pi t^2/s^2)\cdot f_s(\lat) \; \text.
  \]
\end{lemma}

\begin{lemma}%
  \label{lem:roughness-bounds}
  Let $r > r_{0}$ and $H=\spn(1,1)$.
  Then
  \[
    \frac{1}{1 + \eps_{\Z,r/\sqrt{2}}} \; , \;
    \frac{1}{1 + \eps_{\lat_{q},r}(H)} > E(r) = 1 - 2 \exp(-\pi r^{2}/2)
    \; \text.
  \]
\end{lemma}

\begin{proof}
  We first bound $\eps_{\Z,r/\sqrt{2}}$.
  Let $s = r/\sqrt{2} > \sqrt{\ln(2)/\pi}$.
  By the definition of roughness (\cref{def:roughness}) and the facts that $\Z^{*} = \Z$, $\widehat{f_{s}} = s \cdot f_{1/s}$ (\cref{lem:time-scale}), and $f_{1/s}(0)=1$,
  \[ 1 + \eps_{\Z,s}
    = \widehat{f_{s}}(\Z)/\widehat{f_{s}}(0)
    = f_{1/s}(\Z)
    \; \text.
  \]
  Now, by \cref{lem:Banaszczyk} and rearranging (where the denominator is positive due to the lower bound on~$s$),
  \[
    f_{1/s}(\Z) = f_{1/s}(0) + f_{1/s}(\Z \cap T_{1,1})
    < 1 + 2\exp(-\pi s^{2}) \cdot f_{1/s}(\Z)
    \leq \frac{1}{1 - 2\exp(-\pi s^{2})}
    \; \text,
  \]
  which yields the claim.

  For $\eps_{\lat,r}(H)$ where $\lat=\lat_{q}$, again by \cref{def:roughness} and \cref{lem:time-scale},
  \[ 1 + \eps_{\lat,r}(H)
    = \frac{\widehat{f_{r}}(\lat^{*})}{\widehat{f_{r}}(\lat^{*} \cap H^{\perp})}
    = \frac{f_{1/r}(\lat^*)}{f_{1/r}(\lat^{*} \cap H^{\perp})}
    \; \text.
  \]
  Because $H^{\perp}=\spn(1,-1)$ and the vectors $(0,1), (1,-1)/q$ form a basis of~$\lat^{*}$, every point in~$\lat^*$ lies on one of the lines (i.e., affine subspaces) $L_{k} = k\cdot (1,0) + H^{\perp}$ for some $k\in \Z$.
  The unit vector $\vecu = (1,1)/\sqrt{2}$ is orthogonal to~$H^{\perp}$, so for any $\vecx \in L_k$, we have that $\inprod{\vecx}{\vecu}=k/\sqrt{2}$, and hence $\abs{\inprod{\vecx}{\vecu}} \geq 1/\sqrt{2}$ if $k \neq 0$.
  Therefore, $\lat^{*}$ can be partitioned as the disjoint union
  \begin{equation*}
    \lat^* = (\lat^* \cap H^{\perp}) \cup (\lat^* \cap T_{\vecu, 1/\sqrt{2}})
    \; \text.
  \end{equation*}
  So, by \cref{lem:Banaszczyk} and rearranging (where again the denominator is positive due to the bound on~$r$),
  \begin{align*}
    f_{1/r}(\lat^*)
    &= f_{1/r}(\lat^* \cap H^{\perp}) + f_{1/r}(\lat^* \cap T_{\vecu, 1/\sqrt{2}}) \\
    &< f_{1/r}(\lat^* \cap H^{\perp}) + 2\exp(-\pi r^2/2)\cdot f_{1/r}(\lat^*) \\
    &\leq \frac{f_{1/r}(\lat^* \cap H^{\perp})}{1-2\exp(-\pi r^2/2)} .
  \end{align*}
  The result follows by dividing $f_{1/r}(\lat^{*} \cap  H^{\perp})$ by both sides.
\end{proof}

\subsection{Worst-Case Decoding}%
\label{sec:worst-case-ell2}

We now address list-decoding in the~$\ell_{2}$ metric, under worst-case error of bounded distance, by specializing the material of \cref{sec:worst-ell_p} to $p=2$ and using our bounds on $f_{s}(\lat_{q})$ from \cref{sec:bounds-ell2}.
So, we consider decoding distance $d = \delta \sqrt{n}$, where~$n$ is the code length and~$\delta$ is the relative decoding distance.
Then by~\cref{eq:Bpqdelta,eq:rate-versus-distance-ell_p}, we can list-decode for any~$R^{*}$ less than
\begin{equation}
  \label{eq:rate-versus-distance-ell2}
  R^{*,(2)}_{\text{wc},q}(\delta)
  = \sup_{s > 0} W^{(2)}_{q,\delta}\parens{s}^{2}
  > \sup_{s \in (r_{0},q/r_{0})} \frac{\sqrt{2} \cdot \exp(-2\pi \delta^{2}/s^{2})}{s} \cdot E_{q}\parens{s}^{2}
  \; \text,
\end{equation}
where the inequality follows by \cref{lem:common-ell2}.

\Cref{cor:worst-ell2} below is obtained by nearly maximizing the right-hand side of~\eqref{eq:rate-versus-distance-ell2}.
More specifically, a standard calculation shows that taking $s = \delta\sqrt{4\pi}$ maximizes the ``main term'' $\sqrt{2} \cdot \exp(-2\pi\delta^{2}/s^{2})/s$, to have value $1/(\delta\sqrt{2 \pi e})$.
For moderate or larger values of~$\delta$ (and hence~$s$), this very nearly maximizes the entire expression, because $E_{q}\parens{s} \geq E\parens{s}$ since $q/s \geq s$, and $E\parens{s}$ rapidly approaches~$1$ as~$s$ grows.
For example, $E\parens{s}^{2} \geq 1-10^{-8}$ for $\delta \geq 1$.
So, as~$\delta$ grows, the~$R^{*}$ for which we can list-decode rapidly approaches $1/(\delta\sqrt{2 \pi e})$.

\begin{corollary}%
  \label{cor:worst-ell2}
  For any $\delta > \sqrt{\ln(4)}/(2\pi) \approx 0.1874$ and prime $q \geq 4\pi \delta^{2}$, the GS algorithm using weight vector given by~$f_{s}$ for $s = \delta \sqrt{4\pi}$ list-decodes, up to~$\ell_{2}$ distance $\delta\sqrt{n}$ in time $\poly(n,q,1/(\sqrt{\widetilde{R}^{*,(2)}_{\text{wc},q}(\delta)} - \sqrt{R^{*}}))$, any GRS code with adjusted rate
  \[ R^{*} < \widetilde{R}^{*,(2)}_{\text{wc},q}(\delta) \defn \frac{1}{\delta \sqrt{2\pi e}} \cdot E_q\parens{\delta \sqrt{4\pi}}^{2}
    \; \text.
  \]
\end{corollary}

\begin{proof}
  For $s = \delta\sqrt{4\pi}$, the lower bounds on~$\delta$ and~$q$ imply that $s = \delta \sqrt{4\pi} \in (r_{0}, q/r_{0})$.
  Then by hypothesis and \cref{lem:common-ell2,eq:Bpqdelta},
  \begin{equation*}
    R^{*} 
    < \widetilde{R}^{*,(2)}_{\text{wc},q}(\delta)
    = \frac{1}{\delta \sqrt{2\pi e}} \cdot E_{q}\parens{\delta \sqrt{4\pi}}^{2}
    < \frac{\exp(-2\pi \delta^2 / s^2)}{f_s(\lat_q)}
    = W^{(2)}_{q,\delta}\parens{s}^{2}
    \; \text.
  \end{equation*}
  The claim then follows directly by \cref{thm:worst-ell_p}.
\end{proof}

\paragraph*{Comparison to~\cite{DBLP:journals/tit/MookP22}.}

The previous best result for list-decoding (Generalized) Reed--Solomon codes in the~$\ell_2$ metric was given by Mook and Peikert~\cite{DBLP:journals/tit/MookP22}.\footnote{By a standard reduction, the result from~\cite{DBLP:journals/tit/MookP22} also applies to GRS codes, not just RS codes as was originally stated.}

\begin{proposition}[{{\cite[Theorem~3.4]{DBLP:journals/tit/MookP22}}}]%
  \label{prop:MP-dist-bound}
  For any GRS code $\code \subseteq \F_q^n$ with any adjusted rate $R^* < 1$ and any $\eps > 0$, there is a $\poly(n,q,1/\eps)$-time algorithm that list-decodes~$\code$ up to~$\ell_2$ distance $d = \sqrt{n (1-R^*)(1-\eps) / 2}$.
\end{proposition}

Equivalently, for a relative decoding distance $\delta = d/\sqrt{n} > 0$, the result from~\cite{DBLP:journals/tit/MookP22} works for adjusted rates~$R^{*}$ approaching $1-2\delta^{2}$, so it applies only for 
\[ \delta \leq \sqrt{(1-R^{*})/2} \leq 1/\sqrt{2} %
  \; \text.
\]
By contrast, our \cref{thm:worst-ell_p} works for \emph{any} (arbitrarily large) $\delta > 0$ (and \cref{cor:worst-ell2} gives a simpler and more explicit rate bound for any $\delta > 0.1875$).
Moreover, for those~$\delta$ for which both \cref{thm:worst-ell_p,prop:MP-dist-bound} apply, our result works for a larger~$R^{*}$ as long as $R^{*,(2)}_{\text{wc},q}(\delta) > 1-2\delta^{2}$ (see~\eqref{eq:rate-versus-distance-ell_p}).
For typical (moderate or larger)~$q$, this holds for all $\delta \gtrapprox 0.51797$, which corresponds to $R^{*} \lessapprox 0.46342$.
(For tiny $\delta \approx 0$, \cref{thm:worst-ell_p} works for $R^{*} \approx 0.93700$, whereas~\cite{DBLP:journals/tit/MookP22} works for $R^{*} \approx 1$, so the latter is better for very small distances.)

We also point out that~\cite{DBLP:journals/tit/MookP22} proves that for any $\delta \leq 1/2$, which corresponds to $R^* \geq 1/2$, its (very simple) choice of weight vector gives an \emph{optimal} tradeoff between~$\delta$ and~$R^*$ \emph{for the GS/KV soft-decision algorithm and analysis}.
However, the optimality argument breaks down for $\delta > 1/2$ (equivalently, for $R^* < 1/2$).
And indeed, as we have just seen, we obtain a better distance-rate tradeoff than~\cite{DBLP:journals/tit/MookP22} for \emph{almost all} such~$\delta$.
This highlights the interesting question of determining an optimal choice of weights for the GS soft-decision algorithm for $\delta > 1/2$ (especially at the low end of this range).

\subsection{Unique Decoding for a Subclass of GRS Codes}%
\label{sec:ell2-unique}

For a certain natural subclass of GRS codes, and certain rates and decoding distances covered by our list-decoding algorithm, decoding is in fact \emph{unique} (i.e., the list size is at most one).
We show this by giving a lower bound on the~$\ell_{2}$ minimum distance of such codes, and then observing that our list-decoding algorithm can decode to beyond half this distance for all small enough rates.

\begin{lemma}[{{adapted from~\cite[Theorem~4]{DBLP:journals/tit/RothS94}}}]%
  \label{lem:ell2-min-dist}
  Any prime-field GRS code $\grs_{q,k}(\vecalpha,\vecalpha) \subseteq \F_q^n$ (whose twist factors~$\vect$ equal the nonzero evaluation points~$\vecalpha$) of rate $R=k/n$ has \emph{squared} $\ell_2$ minimum distance at least
  \[  \frac{\parens{n+1}^{2} - k^{2}}{12 k^{2}} \cdot (n+1)
    > \frac{1-R^{2}}{12 R^{2}} \cdot n \; \text. \]
\end{lemma}

\begin{proof}
  Any codeword of $\grs_{q,k}(\vecalpha,\vecalpha)$ consists of the evaluations at~$\vecalpha$ of a polynomial of degree at most~$k$ whose constant term is zero.
  Consider any non-zero codeword $\vecc \in \grs_{q,k}(\vecalpha,\vecalpha)$ and let $u(x) \in \F_q[x]$ be its defining polynomial.
  Since $u(x)$ is nonzero with degree at most~$k$, it evaluates to any particular element of $\F_q$ at no more than~$k$ evaluation points.
  Hence, any non-zero element of $\F_q$ appears with multiplicity at most~$k$ in~$\vecc$, and zero appears with multiplicity at most $k-1$, because $u(0)=0$ and zero is not an evaluation point.
  So, $\vecc$ has at most $k-1$ zeros, and at most~$k$ each of $\pm 1, \pm 2, \ldots, \pm \ell$, where $\ell \defn \floor{(n+1-k) / (2k)}$; the remaining $r \defn n+1-k(2\ell+1)$ or more coordinates all have magnitudes greater than~$\ell$.
  Thus,
  \begin{align*}
    \norm{\vecc}_2^2
    &\geq 2k \cdot \sum_{i=1}^{\ell} i^2 + r \parens{\ell+1}^2 \\
    &= (\ell+1) \cdot \parens{k \cdot \ell (2\ell+1)/3 + r(\ell+1)}
      \; \text.
  \end{align*}

  Now define $\beta \defn (n+1-k)/(2k) = \ell + \gamma$, whose fractional part is $\gamma \defn r/(2k) \in [0,1)$, so $r = 2k \cdot \gamma$.
  Then the above bound is
  \[ k (\beta - \gamma + 1) \cdot \parens[\big]{(\beta-\gamma)(2(\beta-\gamma) + 1)/3 + 2 \gamma (\beta - \gamma + 1)} \geq k \cdot \beta (\beta+1) (2\beta+1) / 3 \; \text, \] where the inequality follows from the fact that the minimum over $\gamma \in [0,1)$ is obtained at $\gamma=0$.
  (This can be seen by the closed interval method on $[0,1]$, i.e., differentiating with respect to~$\gamma$, and evaluating at the critical and boundary points.)
  Substituting $\beta (\beta+1) = (\parens{n+1}^{2} - k^{2})/(4k^{2})$ and $k \cdot (2\beta+1) = n+1$, the claim follows.
\end{proof}

\Cref{lem:ell2-min-dist} gives a relationship between the code rate~$R$ and (a lower bound on) half the~$\ell_{2}$ minimum distance, for which decoding to that distance yields a unique solution.
By taking the functional inverse of half this minimum-distance bound, we see that decoding to relative distance~$\delta$ yields a unique solution as long as
\begin{equation}
  \label{eq:Runiq2}
  R < R^{(2)}_{\text{uniq}}(\delta) \defn \frac{1}{\sqrt{48 \delta^{2} + 1}} \; \text,
\end{equation}
which approaches $1/\parens{4 \sqrt{3} \delta}$ as~$\delta$ grows.
This curve is shown in \cref{fig:plots}.
Observe that for any~$\delta$ for which our list-decoding algorithm outperforms the one of~\cite{DBLP:journals/tit/MookP22}, we have that $R^{*,(2)}_{\text{wc}}(\delta) > R^{(2)}_{\text{uniq}}(\delta)$.
In other words, we can efficiently list decode to relative distance~$\delta$ for all rates up to~$R^{(2)}_{\text{uniq}}(\delta)$ (and beyond), thus yielding a \emph{unique} decoder for these parameters.
Alternatively, as the rate~$R$ approaches zero, we can efficiently list decode to a multiple of the unique-decoding distance bound that approaches $4\sqrt{3}/\sqrt{2\pi e} \approx 1.6764$.

\subsection{Average-Case Decoding}%
\label{sec:avg-case-ell2}

We now consider average-case decoding under a memoryless additive (continuous or discrete) Gaussian channel, by specializing the material of \cref{sec:avg-case-ell_p} to $p=2$ and using our bounds on $f_{s}(\lat_{q})$ from \cref{sec:bounds-ell2}.
Consider a Gaussian channel of parameter $r > 0$.
Then by \cref{eq:Apqr,eq:rate-versus-channel-ell_p}, we can list-decode for any $R^{*}$ less than
\begin{equation}
  \label{eq:rate-vs-channel-Gaussian}
  R^{*,(2)}_{\text{ac},q}(r)
  = \sup_{s > 0} A^{(2)}_{q,r}\parens{s}^2 
  > \sup_{s \in (r_{0},q/r_{0})} \frac{s \sqrt{2}}{r^{2}+s^{2}} \cdot E_{q}\parens{s}^{2} 
  \; \text,
\end{equation}
where the inequality is by \cref{lem:common-ell2}.

\Cref{cor:avg-ell2} below is obtained by nearly maximizing the right-hand side of~\eqref{eq:rate-vs-channel-Gaussian}.
More specifically, setting $s=r$ maximizes the ``main term'' $s \sqrt{2}/(r^{2}+s^{2})$, to have value $1/(r \sqrt{2})$.
As above, for moderate or larger values of~$r$ (and hence~$s$), this very nearly maximizes the entire expression, because $E_{q}(s)$ rapidly approaches~$1$ as~$s$ grows.\footnote{By contrast, $E_q(s) \ll 1$ for values of~$s$ very close to~$r_{0}$, in which case the bound is maximized by taking~$s$ somewhat larger than~$r$.}
So, as~$r$ grows, the rate~$R^{*}$ for which we can list-decode rapidly approaches $1/(r\sqrt{2})$.

\begin{corollary}%
  \label{cor:avg-ell2}
  For any $r \in (r_{0}, q/r_{0})$, $\alpha \in (0,1)$, and prime $q$, the GS algorithm using weight vector given by $f_{r}$ list-decodes, in time $\poly(n,q,1/(\sqrt{\widetilde{R}^{*,(2)}_{\text{ac},q}(r)}-\sqrt{R^{*}}))$, any GRS code with adjusted rate
  \[ R^{*} < \widetilde{R}^{*,(2)}_{\text{ac},q}(r) \defn \frac{1}{r\sqrt{2}} \cdot E_q\parens{r}^{2} \; \text,
  \]
  except with probability less than
  $\exp\parens[\big]{-\sqrt{2} n \cdot \alpha^{2} \cdot r \cdot \parens[\big]{\sqrt{\widetilde{R}^{*,(2)}_{\text{ac},q}(r)} - \sqrt{R^{*}}}^{2}}$.
\end{corollary}

\begin{proof}
  By hypothesis, \cref{lem:mu-bound-ell_p,lem:common-ell2,eq:Apqr},
  \begin{equation*}
    R^{*} 
    < \frac{1}{r \sqrt{2}} \cdot E_q\parens{r}^{2}
    < \frac{\mu_{r,r}^{2}}{f_r(\lat_q)}
    = A^{(2)}_{q,r}\parens{r}^{2}
    \; \text.
  \end{equation*}
  The claim then follows directly by \cref{thm:avg-ell_p}, and the fact that $f_{r}(\lat_{q}) > r/\sqrt{2}$ by \cref{lem:fsL-ell2}.
\end{proof}

\section{The $\ell_1$ Metric and Laplacian Error}%
\label{sec:ell1}

In this section, we consider the~$\ell_1$ metric and Laplacian channels.
We specialize \cref{eq:fp} to $p=1$, i.e., the Laplacian function
\[ f(x) \defn f^{(1)}(x) = \exp(-2 \abs{x}) \; \text.
\]
(The Fourier transform of this function is given by
$\widehat{f}(w) = 1/ (1 + \parens{\pi w}^2)$, but we will not use this; as already noted earlier, $f^{(1)}$ satisfies \cref{ass:f-properties}.)

Throughout this section we use the hyperbolic tangent function \[ \tanh(x) \defn \frac{e^{x} - e^{-x}}{e^{x} + e^{-x}}
  = \frac{1-e^{-2x}}{1+e^{-2x}}
  = \frac{e^{2x}-1}{e^{2x}+1}
  < 1
\]
and its reciprocal $\coth(x) = 1/\tanh(x) > 1$.
Observe that $\tanh(x)$ approaches~$1$ as~$x$ grows; it also satisfies $\tanh(x) < x$ for all $x > 0$, and approaches~$x$ as~$x$ approaches zero.\footnote{Both facts can be seen from the Taylor series $\tanh(x) = x - x^{3}/3 + \cdots$, valid for $\abs{x} < \pi/2$.}

\subsection{Bounds}%
\label{sec:bounds-ell1}

In this subsection, we analyze the exact value of $f_s(\lat_q)$ and derive an asymptotic bound.
This appears in the quantities that govern the adjusted rates under which we can decode in the worst and average cases (\cref{eq:Bpqdelta,eq:Apqr}, respectively).
For this purpose, we define a suitable ``fudge factor''.
For any real $x > 0$, define
\begin{equation}
  \label{eq:E-l1}
  E(x) \defn \parens[\Big]{\coth(x) + \frac{4x \cdot e^{2x}}{\parens{e^{2x}-1}^2}}^{-1}
  \in (0,1)
  \; \text,
\end{equation}
where the upper bound comes from the fact that $\coth(x) > 1$.
Note that, as~$x$ grows, the first term in the sum rapidly approaches one, and the second term rapidly approaches zero.
More precisely, a brief calculation reveals that
\begin{equation}
  \label{eq:E-l1-asymptotic}
  E(x) = 1 - O(x \cdot e^{-2x}) \; \text.
\end{equation}

\begin{lemma}%
  \label{lem:common-ell1}
  For any $s > 0$ and positive integer~$q$,
  \[ \frac{1}{f_{s}(\lat_{q})} > \tanh(2/s) \cdot E(q/s) \; \text.
  \]
\end{lemma}

Note that by \cref{eq:E-l1-asymptotic}, for any fixed $s > 0$, as~$q$ (or equivalently, $q/s$) grows, $1/f_{s}(\lat_{q})$ rapidly approaches $\tanh(2/s)$.
In turn, this approaches $2/s$ as~$s$ grows.

\begin{proof}
  This follows directly from \cref{lem:fsL-ell1} below and \cref{eq:E-l1}.
  By \cref{lem:fsL-ell1}, the bound $\coth(2/s) > s/2$, and the definition of $E(x)$,
  \begin{equation*}
    f_s(\lat_{q})
    < \coth(2/s) \parens[\Big]{\coth(q/s) + \frac{2q \cdot e^{2q/s}}{(e^{2q/s} - 1)^2} \cdot \frac{2}{s}}
    = \frac{\coth(2/s)}{E(q/s)}
    \; \text.
  \end{equation*}
  The claim then follows by taking reciprocals.
\end{proof}

\begin{lemma}%
  \label{lem:fsL-ell1}
  For any $s>0$ and positive integer~$q$,
  \begin{equation*}
    f_s(\lat_{q})
    = \coth(2/s) \cdot \coth(q/s) + \frac{2q \cdot e^{2q/s}}{(e^{2q/s} - 1)^2}
    \; \text.
  \end{equation*}
\end{lemma}

\begin{proof}
  By \cref{lem:direct-sum-multiplicative}, we can write
  \begin{equation*}
    f_s(\lat_{q})
    =\sum_{\vecv\in\lat_{q}} f_s(\vecv)
    =\sum_{x=0}^{q-1} f_s((x,x)+q\Z^2)
    =\sum_{x=0}^{q-1} f_s\parens{x+q\Z}^2.
  \end{equation*}
  We first rewrite $f_{s}(x+q\Z)$ as a sum of two geometric series:
  \begin{align*}
    f_s(x + q\Z)
    &= \sum_{z\in\Z} \exp(-2\abs{x+qz}/s)
    \\ &= \sum_{z \geq 0} \exp\parens[\big]{-2\parens{x+qz}/s} + \sum_{z<0} \exp\parens[\big]{2\parens{x+qz}/s}
    \\ &= \sum_{z \geq 0} \exp\parens[\big]{-2\parens{x+qz}/s} + \sum_{z \geq 0} \exp\parens[\big]{2\parens{x-q(z+1)}/s}
    \\ &= \sum_{z \geq 0} \parens[\big]{\exp\parens[\big]{-2(x+qz)/s} + \exp\parens[\big]{-2\parens{q(z+1)-x}/s}}
    \\ &= \frac{e^{-2x/s} + e^{-2(q-x)/s}}{1 - e^{-2q/s}}
    \; \text.
  \end{align*}
  
  Substituting this into the summation, we get that
  \begin{align*}
    \parens{1-e^{-2q/s}}^{2} \cdot f_s(\lat_{q})
    &= \sum_{x=0}^{q-1} \parens{e^{-2x/s} + e^{-2(q-x)/s}}^2
    \\ &= \sum_{x=0}^{q-1} \parens[\big]{e^{-4x/s}+e^{-4(q-x)/s}+2e^{-2q/s}}
    \\ &= \frac{1 - e^{-4q/s}}{1 - e^{-4/s}} \cdot \parens{1 + e^{-4/s}} + 2q \cdot e^{-2q/s}
    \\ &= \coth(2/s) \cdot (1 - e^{-4q/s}) + 2q \cdot e^{-2q/s}
    \\ &= \coth(2/s) \cdot \frac{1+e^{-2q/s}}{1-e^{-2q/s}} \cdot \parens{1-e^{-2q/s}}^{2} + 2q \cdot e^{-2q/s}
    \\ &= \coth(2/s) \cdot \coth(q/s) \cdot \parens{1-e^{-2q/s}}^{2} + 2q \cdot e^{-2q/s}
         \; \text.
  \end{align*}
  The claim follows by dividing both sides by $\parens{1-e^{-2q/s}}^{2}$, and multiplying both the numerator and denominator of the final term by $e^{4q/s}$.
\end{proof}

\subsection{Worst-Case Decoding}%
\label{sec:worst-case-ell1}

Now we address list-decoding in the~$\ell_{1}$ metric, under worst-case error of bounded distance, by specializing the material of \cref{sec:worst-ell_p} to $p=1$ and using our bound on $f_s(\lat_q)$ from \cref{lem:common-ell1}.
We consider decoding distance $d = \delta n$, where~$n$ is the code length and~$\delta$ is the relative decoding distance.
Then by \cref{eq:Bpqdelta,eq:rate-versus-distance-ell_p,lem:common-ell1}, we can list-decode for any $R^*$ less than
\begin{equation}
  \label{eq:rate-versus-distance-ell1}
  R^{*,(1)}_{\text{wc},q}(\delta)
  = \sup_{s > 0} W^{(1)}_{q,\delta}\parens{s}^{2}
  > \sup_{s > 0} \exp(-4 \delta/s) \cdot \tanh(2/s) \cdot E(q/s)
  \; \text.
\end{equation}

\Cref{cor:worst-ell1} below is obtained by maximizing the ``main term'' $\exp(-4\delta/s) \cdot \tanh(2/s)$ of the right-hand side of~\eqref{eq:rate-versus-distance-ell1}.
By calculus, this is done by taking $s = 4/\ln(D(\delta)) > 0$, where
\begin{equation*}
  \label{eq:D-delta}
  D(\delta) \defn \sqrt{1 + \frac{1}{\delta^{2}}} + \frac{1}{\delta} > 1 \; \text.
\end{equation*}
Substituting, this means we can list-decode for any~$R^{*}$ less than
\begin{equation}
  \label{eq:Rtilde-star1}
  \widetilde{R}^{*,(1)}_{\text{wc},q}(\delta)
  \defn \frac{\tanh(\ln \sqrt{D(\delta)})}{D\parens{\delta}^{\delta}} \cdot E(q \ln(D(\delta))/4)
  = \frac{D(\delta)-1}{D(\delta)+1} \cdot \frac{E(q \ln(D(\delta))/4)}{D\parens{\delta}^{\delta}}
  \; \text.
\end{equation}
We consider this quantity's asymptotic behavior for large and small~$\delta$:
\begin{itemize}
\item As~$\delta$ grows, $D(\delta) = 1 + 1/\delta + O(1/\delta^{2})$ and $D\parens{\delta}^{\delta}$ approaches~$e$, hence $\widetilde{R}^{*,(1)}_{\text{wc},q}(\delta)$ approaches $1/(2e \delta)$ as $q/\delta$ also grows.
  This is consistent with \cref{rem:optimize-ell_p}.
\item As~$\delta$ approaches zero, $D(\delta)$ approaches $2/\delta$ and $D\parens{\delta}^{\delta}$ approaches~$1$, hence $\widetilde{R}^{*,(1)}_{\text{wc},q}(\delta)$ approaches~$1$ as $q/\delta$ also grows.
\end{itemize}

Alternatively, we can get a simpler but cruder bound by replacing $\tanh(2/s)$ in \cref{eq:rate-versus-distance-ell1} with its upper bound of $2/s$.
Then the resulting ``main term'' of $2 \exp(-4\delta/s)/s$ is maximized at $s = 4\delta$; substituting, this means we can list decode for any~$R^{*}$ less than
\[ e^{-1} \cdot \tanh(1/(2\delta)) \cdot E(q/(4\delta)) \; \text.
\]
This bound approaches $1/(2e \delta)$ as~$\delta$ and $q/\delta$ grow, which matches the behavior of $\widetilde{R}^{*,(1)}_{\text{wc},q}(\delta)$ as described above.
However, as~$\delta$ approaches zero (and $q/\delta$ grows), the above bound merely approaches $1/e$, which is much worse than the limit of~$1$ for $\widetilde{R}^{*,(1)}_{\text{wc},q}(\delta)$.

\begin{corollary}%
  \label{cor:worst-ell1}
  For any $\delta > 0$ and prime~$q$, the GS algorithm using weight vector $f_{s}$ for $s=4/\ln(D(\delta))$ list-decodes, up to~$\ell_{1}$ distance $\delta n$ in time $\poly(n,q,1/(\sqrt{\widetilde{R}^{*,(1)}_{\text{wc},q}(\delta)} - \sqrt{R^{*}}))$, any GRS code with adjusted rate $R^{*} < \widetilde{R}^{*,(1)}_{\text{wc},q}(\delta)$ (see \cref{eq:Rtilde-star1}).
\end{corollary}

\begin{proof}
  By hypothesis and \cref{lem:common-ell1,eq:Bpqdelta},
  \begin{equation*}
    R^{*}
    < \widetilde{R}^{*,(1)}_{\text{wc},q}(\delta)
    = \frac{\tanh(\ln \sqrt{D(\delta)})}{D\parens{\delta}^{\delta}} \cdot E(q/s)
    < \frac{\exp(-4 \delta / s)}{f_s(\lat_q)}
    = W^{(1)}_{q,\delta}\parens{s}^{2}
    \; \text.
  \end{equation*}
  The claim then follows directly by \cref{thm:worst-ell_p}.
\end{proof}

\paragraph*{Comparison to~\cite{DBLP:journals/tit/RothS94,wu03:_lee_bch_reed_solomon}.}

To our knowledge, the only prior algorithms for (unique or list) decoding Reed--Solomon codes in the~$\ell_1$ (Lee) metric are~\cite[Section~5]{DBLP:journals/tit/RothS94} and~\cite{wu03:_lee_bch_reed_solomon}.
We note that both of these require \emph{discrete} (integer) error, whereas our algorithm works for \emph{continuous} error.

For a certain subclass of GRS codes (and BCH codes more generally),~\cite{DBLP:journals/tit/RothS94} gives a \emph{unique} decoding algorithm for up to half (a lower bound on) the~$\ell_{1}$ minimum distance, using Euclid's algorithm for polynomials.
This algorithm decodes up to any relative distance $\delta < 1 - R < 1 - R^{*}$.
For any prime-field GRS code,~\cite{wu03:_lee_bch_reed_solomon} gives a list-decoding algorithm that uses GS as a subroutine, and has a piecewise distance-rate tradeoff due to its optimization over an integer parameter.
(The algorithm works by putting equal weight on a range of alphabet symbols centered at the received symbol, optimizing over the range size for a given rate.)

By contrast with~\cite{DBLP:journals/tit/RothS94}, and like~\cite{wu03:_lee_bch_reed_solomon}, our \cref{cor:worst-ell1} works for \emph{any} GRS code, and for \emph{any} (arbitrarily large) relative decoding distance $\delta > 0$, for sufficiently small $R^{*} > 0$.
Our rate-distance trade-off surpasses that of both~\cite{DBLP:journals/tit/RothS94,wu03:_lee_bch_reed_solomon} for all $\delta \gtrapprox 0.78988$, which corresponds to rates $R^* \lessapprox 0.21012$; see \cref{fig:plots}.

\subsection{Unique Decoding for a Subclass of GRS Codes}%
\label{sec:ell1-unique}

As in \cref{sec:ell2-unique}, for the same subclass of GRS codes and certain parameters covered by our list-decoding algorithm, the decoding output is in fact \emph{unique}.
To show this, we give a lower bound on the~$\ell_{1}$ minimum distance of such codes, and then observe that our list-decoding algorithm can decode to beyond half this distance for all small enough rates.

\begin{lemma}[{{adapted from~\cite[Theorem~4]{DBLP:journals/tit/RothS94}}}]%
  \label{lem:ell1-min-dist}
  Any prime-field GRS code $\grs_{q,k}(\vecalpha,\vecalpha) \subseteq \F_q^n$ (whose twist factors~$\vect$ equal the nonzero evaluation points~$\vecalpha$) of rate $R=k/n$ has $\ell_1$ minimum distance at least
  \[ \frac{\parens{n+1}^2 - k^2}{4k}
    > \frac{1 - R^{2}}{4R} \cdot n \; \text. \]
\end{lemma}

\begin{proof}
  Define $\beta \defn (n+1-k) / (2k)$ and write it in terms of its integer and fractional parts as $\beta = \ell + \gamma$, where $\ell \defn \floor{(n+1-k) / (2k)}$ and $\gamma \defn r / (2k)$ for $r \defn n+1-k(2\ell+1)$.
  Following the same reasoning as in the proof of \cref{lem:ell2-min-dist}, and using the fact that $\gamma < 1$, the~$\ell_1$ distance of any nonzero $\vecc \in \grs_{q,k}(\vecalpha,\vecalpha)$ is
  \begin{align*}
    \norm{\vecc}_1 
    &\geq 2k \cdot \sum_{i=1}^{\ell}i + r(\ell+1) \\
    &= (k\ell + r)(\ell+1) \\
    &= (k(\beta - \gamma) + 2k \gamma)(\beta - \gamma +1) \\
    &= k(\beta + \gamma)(\beta - \gamma +1) \\
    &= k(\beta^2 - \gamma^2 + \beta + \gamma) \\
    &\geq k \beta (\beta+1) \\
    &= \frac{\parens{n+1}^2 - k^2}{4k} \; \text.
  \end{align*}
\end{proof}

\Cref{lem:ell1-min-dist} gives a relationship between the code rate~$R$ and (a lower bound on) half the~$\ell_{1}$ minimum distance, for which decoding to that distance yields a unique solution.
By taking the functional inverse of half this minimum-distance bound, we see that decoding to relative distance~$\delta$ yields a unique solution as long as
\begin{equation}
  \label{eq:Runiq1}
  R < R^{(1)}_{\text{uniq}}(\delta) \defn -4\delta + \sqrt{\parens{4\delta}^{2}+1} \; \text,
\end{equation}
which approaches $1/(8\delta)$ as~$\delta$ grows.
This curve is shown in \cref{fig:plots}.
Observe that for any~$\delta$ for which our list-decoding algorithm outperforms the unique decoder of~\cite{DBLP:journals/tit/RothS94} (or for which~\cite{DBLP:journals/tit/RothS94} does not apply), we have that $R^{*,(1)}_{\text{wc}}(\delta) > R^{(1)}_{\text{uniq}}(\delta)$.
In other words, we can efficiently list decode to relative distance~$\delta$ for all rates up to~$R^{(1)}_{\text{uniq}}(\delta)$ (and beyond), thus yielding a \emph{unique} decoder for these parameters.
Alternatively, as the rate~$R$ approaches zero, we can efficiently list decode to a multiple of the unique-decoding distance bound that approaches $8/(2e) \approx 1.4715$.

\subsection{Average-Case Decoding}%
\label{sec:avg-case-ell1}

We now consider average-case decoding under a memoryless additive (continuous or discrete) Laplacian channel, by specializing the material of \cref{sec:avg-case-ell_p} to $p=1$ and using our bound on $f_s(\lat_q)$ from \cref{lem:common-ell1}.
Consider a Laplacian channel of parameter $r > 0$.
Then by \cref{eq:Apqr,eq:rate-versus-channel-ell_p}, we can list-decode for any $R^*$ less than
\begin{equation}
  \label{eq:rate-vs-channel-Laplacian}
  R^{*,(1)}_{\text{ac},q}(r)
  = \sup_{s > 0} A^{(1)}_{q,r}\parens{s}^2
  > \sup_{s > 0} \frac{s^2 \cdot \tanh(2/s)}{\parens{r+s}^2} \cdot E\parens{q/s} \; \text,
\end{equation}
where the inequality is by \cref{lem:common-ell1}.

\Cref{cor:avg-ell1} below is obtained by nearly maximizing the right-hand side of~\eqref{eq:rate-vs-channel-Laplacian}, at least for moderate or large values of~$r$.
Specifically, we use the bound $\tanh(2/s) < 2/s$ to approximate the ``main term'' of~\eqref{eq:rate-vs-channel-Laplacian} by $2s / \parens{r+s}^2$.
This is maximized at $s=r$, which makes the original main term equal to $\tanh(2/r)/4$.
Note that $R^{*,(1)}_{\text{ac},q}(r)$ does indeed approach this value as~$r$ and $q/r$ grow, because $\tanh(2/r)$ approaches $2/r$, and $E(q/r)$ rapidly approaches~$1$ (see \cref{eq:E-l1-asymptotic}).

However, for small values of~$r$, the expression in~\eqref{eq:rate-vs-channel-Laplacian} is maximized for~$s$ significantly larger than~$r$, to have value much larger than $\tanh(2/r)/4 < 1/4$.
This maximization can be computed numerically, and indeed, $R^{*,(1)}_{\text{ac},q}(r)$ approaches~$1$ as $r$ approaches~$0$; see \cref{fig:plots}.

\begin{corollary}%
  \label{cor:avg-ell1}
  For any $r > 0$, $\alpha \in (0,1)$, and prime $q$, the GS algorithm using weight vector given by $f_r$ list-decodes, in time $\poly(n, q, 1/\parens{\sqrt{\widetilde{R}^{*,(1)}_{\text{ac},q}(r)}-\sqrt{R^*}})$, any GRS code with adjusted rate
  \[ R^{*} < \widetilde{R}^{*,(1)}_{\text{ac},q}(r)
    \defn \frac{\tanh(2/r)}{4} \cdot E\parens{q/r} \; \text,
  \]
  except with probability less than
  $\exp\parens[\big]{-n \cdot \alpha^{2} \cdot r \cdot \parens[\big]{\sqrt{\widetilde{R}^{*,(1)}_{\text{ac},q}(r)} - \sqrt{R^{*}}}^{2}}$.
\end{corollary}

\begin{proof}
  By hypothesis, \cref{lem:mu-bound-ell_p,lem:common-ell1,eq:Apqr},
  \begin{equation*}
    R^{*}
    < \widetilde{R}^{*,(1)}_{\text{ac},q}(r)
    = \frac{\tanh(2/r)}{4} \cdot E\parens{q/r}
    < \frac{\mu_{r,r}^{2}}{f_r(\lat_q)}
    = A^{(1)}_{q,r}\parens{r}^{2}
    \; \text.
  \end{equation*}
  The claim then follows directly by \cref{thm:avg-ell_p}, and (for the probability bound) the fact that $f_{r}(\lat_{q}) > \coth(2/r) > r/2$ by \cref{lem:fsL-ell1}.
\end{proof}

\bibliographystyle{alphaabbrvprelim}

\end{document}